\documentclass[10pt,aps,prc,floatfix,twocolumn,nofootinbib,superscriptaddress]{revtex4-1}
\usepackage{graphicx,amsmath,amssymb,bm}
\usepackage{amsfonts}
\usepackage[utf8]{inputenc}
\usepackage{verbatim}
\usepackage{float}
\usepackage[labelfont={small},font=small,subrefformat=parens,caption=false]{subfig}
\usepackage{cancel}
\usepackage{multirow}
\usepackage{array}
\usepackage{xparse}
\usepackage{xspace}

\usepackage{bigstrut}

\usepackage{physics}
\usepackage{color}
\usepackage{dsfont}

\usepackage{dcolumn}

\usepackage[pdfencoding=auto, pdfpagelabels]{hyperref}

\usepackage[overload]{textcase}

\definecolor{linkcolor}{rgb}{0,0,0.40} 
\hypersetup{%
    pdfsubject=Paper,
    pdfkeywords={nuclear physics} {Bayesian} {chiral EFT} {three body force},
    unicode = true,
    breaklinks = true,
    colorlinks = true,
    linkcolor = linkcolor,
    citecolor = linkcolor,
    menucolor = linkcolor,
    urlcolor = linkcolor
}

\usepackage{cellspace}
\graphicspath{{./figures/}}

\makeatletter
\newcommand\newsubcommand[3]{\newcommand#1{#2\sc@sub{#3}}}
\def\sc@sub#1{\def\sc@thesub{#1}\@ifnextchar_{\sc@mergesubs}{_{\sc@thesub}}}
\def\sc@mergesubs_#1{_{\sc@thesub#1}}

\newcommand\newsupcommand[3]{\newcommand#1{#2\sc@sup{#3}}}
\def\sc@sup#1{\def\sc@thesup{#1}\@ifnextchar^{\sc@mergesups}{^{\sc@thesup}}}
\def\sc@mergesups^#1{^{\sc@thesup#1}}
\makeatother

\DeclareMathAlphabet{\mathbcal}{OMS}{cmsy}{b}{n}

\DeclareMathOperator{\invchisq}{\chi^{--2}}  

\newcommand{\cbar}{\bar c}

\newcommand{\etal}{\textit{et~al.}\xspace}

\newcommand{\NN}{\ensuremath{NN}}

\newcommand{\ordervec}{\vec}

\newcommand{\inputvec}{\mathbf}

\newsubcommand{\ckvec}{\ordervec{c}}{k}

\newsubcommand{\bkvec}{\ordervec{b}}{k}

\newsubcommand{\ckvecset}{\ordervec{\inputvec{c}}}{k}

\newsubcommand{\ckvecapprox}{\mathbf{c}'}{k}
\newsubcommand{\ckvecapproxset}{\mathbf{C}'}{k}

\newsubcommand{\bkvecapprox}{\mathbf{b}'}{k}
\newsubcommand{\bkvecset}{\mathbf{B}}{k}
\newsubcommand{\bkvecapproxset}{\mathbf{B}'}{k}

\newcommand{\genobs}{y}

\newsubcommand{\genobsvec}{\ordervec{\genobs}}{k}
\newsubcommand{\genobsvecset}{\ordervec{\inputvec{\genobs}}}{k}

\newcommand{\genobsexp}{\genobs_{\textup{exp}}}        
\newcommand{\genobsexpset}{\inputvec{\genobs}_{\textup{exp}}}
\newcommand{\genobsth}{\genobs_{\textup{th}}}          
\newcommand{\genobsthset}{\inputvec{\genobs}_{\textup{th}}}

\newcommand{\lecs}{\vec{a}}

\newcommand{\covarth}{{\Sigma}_{\mathrm{th}}}  

\newsubcommand{\akvec}{\mathbf{a}}{k}

\newsubcommand{\akvecapprox}{\mathbf{a}'}{k}
\newsubcommand{\akvecset}{\mathbf{A}}{k}
\newsubcommand{\akvecapproxset}{\mathbf{A}'}{k}

{}  

\DeclareMathOperator{\pr}{pr} 
\newcommand{\given}{\,|\,}  

\newcommand{\abar}{\bar{a}}

\newcommand{\normal}{\mathcal{N}}

\newcommand{\chiEFT}{$\chi$EFT}

\newcommand{\genobsref}{\ensuremath{y_{\mathrm{ref}}}}
\newcommand{\genobsrefset}{\ensuremath{\inputvec{y}_{\mathrm{ref}}}}

\def\diffd{\mathrm{d}}  

\DeclareDocumentCommand\differential{ o g d() }{ 
    \IfNoValueTF{#2}{
        \IfNoValueTF{#3}
            {\diffd\IfNoValueTF{#1}{}{^{#1}}}
            {\mathinner{\diffd\IfNoValueTF{#1}{}{^{#1}}\argopen(#3\argclose)}}
        }
        {\mathinner{\diffd\IfNoValueTF{#1}{}{^{#1}}#2} \IfNoValueTF{#3}{}{(#3)}}
    }
\DeclareDocumentCommand\dd{}{\differential} 

\newcommand{\pathd}{\mathcal{D}}  

\DeclareDocumentCommand\pathdifferential{ o g d() }{ 
    \IfNoValueTF{#2}{
        \IfNoValueTF{#3}
            {\pathd\IfNoValueTF{#1}{}{^{#1}}}
            {\mathinner{\pathd\IfNoValueTF{#1}{}{^{#1}}\argopen(#3\argclose)}}
        }
        {\mathinner{\pathd\IfNoValueTF{#1}{}{^{#1}}#2} \IfNoValueTF{#3}{}{(#3)}}
    }

\newcommand{\ritzbasis}{X}
\newcommand{\coeff}{\beta}

\newcommand{\normmat}{\mathcal{N}}

\newcommand{\EC}{EC}
\newcommand{\NEC}{\ensuremath{N_{\rm EC}}}
\newcommand{\subspace}[1]{ \ensuremath{\widetilde{#1}} }

\newcommand{\cD}{c_D}
\newcommand{\cE}{c_E}
\renewcommand{\NN}{\ensuremath{\mathrm{NN}}}
\newcommand{\piN}{\ensuremath{\pi\mathrm{N}}}
\newcommand{\NNN}{\ensuremath{\mathrm{3N}}}
\newcommand{\TNF}{3NF\xspace}
\newcommand{\TNFs}{\TNF{s}\xspace}

\newcommand{\eg}{\textit{e.g.}\xspace}
\newcommand{\ie}{\textit{i.e.}\xspace}
\renewcommand{\etal}{\textit{et al}\xspace}  

 \newcommand{\verifyvalue}[1]{#1}

\begin{document}

\title{Rigorous constraints on three-nucleon forces in chiral effective field theory\\ from fast and accurate calculations of few-body observables}

\author{S.~Wesolowski}
\email{scwesolowski@salisbury.edu}
\affiliation{Department of Mathematics and Computer Science, Salisbury University, Salisbury, MD 21801, USA}

\author{I. Svensson}
\email{isak.svensson@chalmers.se}
\affiliation{Department of Physics, Chalmers University of Technology, SE-412 96 G\"oteborg, Sweden}

\author{A. Ekstr\"om}
\email{andreas.ekstrom@chalmers.se}
\affiliation{Department of Physics, Chalmers University of Technology, SE-412 96 G\"oteborg, Sweden}

\author{C. Forss\'en}
\email{christian.forssen@chalmers.se}
\affiliation{Department of Physics, Chalmers University of Technology, SE-412 96 G\"oteborg, Sweden}

\author{R.~J. Furnstahl}
\email{furnstahl.1@osu.edu}
\affiliation{Department of Physics, The Ohio State University, Columbus, OH 43210, USA}

\author{J.~A. Melendez}
\email{melendez.27@osu.edu}
\affiliation{Department of Physics, The Ohio State University, Columbus, OH 43210, USA}

\author{D.~R. Phillips}
\email{phillid1@ohio.edu}
\affiliation{Department of Physics and Astronomy and Institute of Nuclear and Particle Physics, Ohio University, Athens, OH 45701, USA}
\affiliation{Institut f\"ur Kernphysik, Technische Universit\"at Darmstadt, 64289 Darmstadt, Germany}
\affiliation{ExtreMe Matter Institute EMMI, GSI Helmholtzzentrum f{\"u}r Schwerionenforschung GmbH, 64291 Darmstadt, Germany}

\date{\today}

\begin{abstract}
   We explore the constraints on the three-nucleon force (3NF) of chiral effective field theory ($\chi$EFT) that are  provided by bound-state observables in the $A=3$ and $A=4$ sectors.
   Our statistically rigorous analysis incorporates experimental error, computational method uncertainty, and the uncertainty due to truncation of the $\chi$EFT expansion at next-to-next-to-leading order.
   A consistent solution for the ${}^3$H binding energy, the ${}^4$He binding energy and radius, and the ${}^3$H $\beta$-decay rate can only be obtained if $\chi$EFT truncation errors are included in the analysis.
 The $\beta$-decay rate is the only one of these that yields a non-degenerate constraint on the 3NF low-energy constants, which makes it crucial for the parameter estimation.
   We use eigenvector continuation for fast and accurate emulation of No-Core Shell Model calculations of 
   the few-nucleon observables.
   This facilitates sampling of the posterior probability distribution, allowing us to also determine the distributions of the parameters that quantify the truncation error.
   We find a $\chi$EFT expansion parameter of $Q=0.33 \pm 0.06$ for these observables.  
\end{abstract}

\maketitle


\section{Motivation and goals} \label{sec:intro}

In low-energy effective field theories (EFTs) of many-body systems, three- and higher-body forces inevitably arise because they capture the effect of degrees of freedom not resolved in the EFT~\cite{Bedaque:1998kg,Hammer:2012id,Capel:2020obz}. 
In the variant of chiral EFT (\chiEFT) without an explicit Delta resonance, three-nucleon forces (\TNFs) first appear in the Hamiltonian at third order (next-to-next-to-leading order) in the EFT expansion. 
This first contribution depends on two parameters, called $\cD$ and $\cE$, not already determined by nucleon-nucleon (\NN) or pion-nucleon (\piN) scattering. The terms proportional to $\cD$ and $\cE$, together with the venerable Fujita-Miyazawa term~\cite{Fujita:1957zz}, form the dominant piece of the \TNF in \chiEFT~\cite{VanKolck:1994yi,Epelbaum:2002vt}. This \TNF has small, but important, effects in light nuclei and helps drive saturation in heavier systems and symmetric nuclear matter~\cite{Hebeler:2020ocj}. But---as in any EFT---$\cD$ and $\cE$ must be estimated from data, either using experimental measurements or theoretical sources. Doing that reliably, with error bars that account for all uncertainties, is key to accurate use of \chiEFT\ forces in computations of nuclei. 

In this work, 
we carry out parameter estimation for $\cD$ and $\cE$ within a Bayesian framework. We explore the constraints on $\cD$ and $\cE$ provided by several observables: the triton and ${}^4$He particle binding energies, the ${}^4$He particle charge radius, and the Gamow-Teller matrix element of the triton, as extracted from tritium $\beta$-decay.
In addition to the standard treatment of uncertainties in the experimental measurements,  we also account for model discrepancy~\cite{Kennedy:2001,Brynjarsdottir:2014} by considering 
the uncertainty in the \chiEFT\  Hamiltonian itself. In particular, we include \chiEFT\ truncation errors in the parameter estimation using a statistical model applied previously in the \NN{} sector~\cite{Wesolowski:2018lzj,Melendez:2017phj}. A novel feature of our analysis is that we employ eigenvector continuation (\EC)~\cite{Frame:2017fah} to implement rapid sampling~\cite{Konig:2019adq,Ekstrom:2019lss} of a multi-dimensional posterior, and hence obtain joint probability distributions for $\cD$, $\cE$, and the EFT expansion parameter, $Q$.
The fits of the \NN{} and \piN{} parameters that are inputs to our calculations also have uncertainties; we propagate the uncertainties from \NN{} but not from \piN{} (see Sec.~\ref{sec:optimization}).  
The outputs from the parameter estimation are not single values for $\cD$ and $\cE$ but multi-dimensional 
posterior probability density functions (pdfs). These---referred to as ``posteriors'' hereafter---can be used to identify correlations and to propagate uncertainties to observables.

This is not an exhaustive study of parameter estimation for these \TNF{} parameters.
Rather our goal is to examine the implications of using particular combinations of observables for constraining $\cD$ and $\cE$ while exemplifying statistical best practices~\cite{Wesolowski:2018lzj}, in particular the inclusion of EFT truncation errors as a guard against overfitting.
There are several recent and ongoing efforts seeking analogous constraints, which can provide complementary information, and many of our conclusions reinforce those of other authors. In particular, 
we build on the use of tritium $\beta$-decay in Refs.~\cite{Gazit:2009PRL, Baroni:2016aa} (cf. Ref.~\cite{Baroni:2018fdn} for an analysis in \chiEFT\ with explicit $\Delta(1232)$ degrees of freedom) and compare our results to the $\cD$--$\cE$ posteriors found using other observables such as Nd scattering~\cite{Epelbaum:2018ogq} and neutron-$\alpha$ scattering~\cite{Kravvaris:2020lhp}.

In Sec.~\ref{sec:strategy} we describe our Bayesian strategy for estimating $\cD$ and $\cE$: our choice of likelihood and prior distributions, including our optimization of the input \NN{} force. Then in Sec.~\ref{sec:fewbodydetails} we discuss details of the few-body methods used to compute observables and introduce the \EC{}
emulators that make our comprehensive parameter-estimation process feasible. 
Results are given in Sec.~\ref{sec:results}, first for the most comprehensive fit of the \TNF\ parameters and then using constraints provided by individual observables.
We identify the induced correlations, infer knowledge of the EFT expansion, and display the range of \chiEFT\ predictions obtained from our $\cD$ and $\cE$ posterior. 
Our takeaway points and avenues for future work are summarized in Sec.~\ref{sec:summary}. An open-source python package \texttt{fit3bf} accompanies this article~\cite{fit3bf} and can be used to reproduce all the figures herein.


\section{Bayesian Strategy} \label{sec:strategy}

Our aim is to determine \TNF low-energy constants (LECs) $\{\cD, \cE\}$ from experimental data $\genobsexpset$. The few-body observables in $\genobsexpset$ 
are the mass and radius of ${}^4$He, and the mass and $\beta$-decay rate of ${}^3$H. 
The Bayesian approach we implement can account for all sources of uncertainty: from data, from the theoretical model, and from the calculational methods~\cite{Wesolowski:2015fqa,Wesolowski:2018lzj}.
Some of these will not be treated in this work because they are either negligible (\eg, emulator error; see Sec.~\ref{sec:emulator}) or more work needs to be done to properly include them ($\piN{}$ LECs; see Sec.~\ref{sec:optimization}).
The largest source of uncertainty is the \chiEFT\ truncation error, but we also account for the experimental and the few-nucleon solver uncertainties.
Our use of emulators makes the observable calculations required for Markov chain Monte Carlo (MCMC) sampling rapid enough that we can fully account for $\NN{}$ uncertainties and incorporate truncation uncertainty in a Bayesian fashion. 

In this section we first detail our approach to assessing truncation errors~\cite{Furnstahl:2015rha,Melendez:2019izc}. We then write down the forms for the posterior and prior, before describing how the convergence pattern
of $A=3$ and $A=4$ observables provide information on the truncation error. 
The section closes with a description of how the \NN{} LEC values and uncertainties that are input to our calculation are obtained.
 
\subsection{Including EFT truncation error} 

We follow a Bayesian approach for the consistent incorporation of all higher-order terms in the EFT~\cite{Melendez:2019izc}.
Let $\genobsth(\lecs)$ be the prediction of some observable $\genobs$ at a fixed order in the EFT and for fixed values of LECs $\lecs$. 
Here, $\lecs$ includes the $\NN{}$ LECs along with $\cD$ and $\cE$.
Dependence on the $\piN{}$ LECs is left implicit throughout; see Sec.~\ref{sec:LECpriors}.
We account for the presence of  theory and experimental uncertainties $\delta\genobsth$ and $\delta\genobsexp$ by writing~\cite{Kennedy:2001,Brynjarsdottir:2014,Wesolowski:2018lzj}:
\begin{align} \label{eq:relation_exp_and_th}
    \genobsexp = \genobsth(\lecs) + \delta \genobsth + \delta \genobsexp
    \,.
\end{align}
That is, the theoretical value differs from the measured value because of {\it both} experimental uncertainties and discrepancies in the theory.
For the measurement errors $\delta \genobsexp$ we assume a Gaussian error term that is uncorrelated between observables. However, this assumption has little impact on our results because experimental errors are small relative to theory uncertainties.

The distribution of the theory discrepancy $\delta\genobsth$ also follows a Gaussian distribution~\cite{Furnstahl:2015rha}.
It depends on two dimensionless parameters related to the EFT convergence pattern.
The first is the EFT expansion parameter $Q$, which is a number in $(0, 1)$ and governs the factor by which each correction should shrink in a well constructed EFT\@.
The model encodes the expectation that the first omitted term in a \chiEFT\ of order $k$ is of order $\genobsref \cbar\, Q^{k+1}$, where $\genobsref$ is the known characteristic size of the observable $\genobs$~\cite{Furnstahl:2015rha, Melendez:2017phj}.
The second dimensionless parameter is then $\cbar$.
It governs the magnitude of the relative correction at each order after we have accounted for $Q$.

For a given $\cbar$ and $Q$ the error due to all terms beyond $O(Q^k)$ in the EFT can be summed and used to create a covariance matrix between observable $i$ and observable $j$. In this work we assume that there are no correlations between the EFT errors for the observables of interest, thus the covariance matrix is diagonal~\cite{Wesolowski:2018lzj}:
\begin{align}
    (\covarth)_{ij} & = \left[\frac{(\genobsref \cbar\, Q^{k+1})^2}{1 - Q^2}\right]\delta_{ij} \;. 
    \label{eq:theorycovar}
\end{align}
We view this as the simplest form of $\covarth$ that models the effect of higher-order terms in the \chiEFT\ expansion. There are certainly other plausible forms of $\covarth$ that invoke correlated EFT uncertainties, \eg, we could assume the fourth-and-higher order contributions to these observables are correlated according to the pattern of correlations observed between them at lower orders, cf.\  Refs.~\cite{Melendez:2019izc,Drischler:2020yad,Maris:2020qne}. As we mention in Sec.~\ref{sec:summary} below, exploring the impact of more sophisticated forms of $\covarth$ on the results is an avenue for future work. Practitioners who wish to examine such possibilities themselves should find it straightforward to do so using the open-source python package \texttt{fit3bf} that accompanies this article~\cite{fit3bf}.

\subsection{The pdf for \texorpdfstring{$\cD$}{cD} and \texorpdfstring{$\cE$}{cE}}

The form of the experimental and theory uncertainties and the relation~\eqref{eq:relation_exp_and_th} are sufficient to determine that the likelihood is given by:
\begin{equation} \label{eq:likelihood}
      \pr(\genobsexpset \given \lecs, \Sigma, I)  \sim \normal[\genobsthset, \Sigma].
\end{equation}
This likelihood is a multivariate Gaussian pdf, defined by central values from theory, $\genobsthset$, and the covariance matrix $\Sigma \equiv \Sigma_{\rm exp}+\Sigma_{\rm method}+\Sigma_{\rm th}$, where we have also included a term $\Sigma_{\rm method}$ that describes the uncertainty of our few-nucleon solver.
Here, the combination $\Sigma_{\rm exp}+\Sigma_{\rm method}$ is a diagonal matrix given by the column of adopted errors in Table~\ref{tab:orderbyorder}.
The precision of our few-nucleon calculations is discussed in Sec.~\ref{sec:fewbodydetails}.
The covariance matrix could be extended to include a term from the emulators, but we do not do that here as those errors are negligible. 

If the truncation error parameters $\cbar$ and $Q$ appearing in $\Sigma_{\rm th}$ are known from prior information then this likelihood, together with priors on $\lecs$, defines the posterior probability density to be computed.
Although there is some evidence that suggests $Q \approx 0.3$~\cite{Binder:2018pgl,Epelbaum:2019kcf,Maris:2020qne}, we use uninformative assumptions so as not to bias our results unnecessarily.
We handle this by treating $\cbar$ and $Q$ as additional random variables; that is, we assign priors to them and learn their posterior distributions in tandem with the LECs.

The full joint pdf for all these parameters of interest then follows from Bayes' theorem:
\begin{align} \label{eq:master_posterior}
\begin{split}
    \pr(\lecs, \cbar^2, Q \given \genobsexpset, I) & \propto \pr(\genobsexpset \given \lecs, \Sigma, I) \pr(\lecs \given I) \\
    & \times \pr(\cbar^2 \given Q, \lecs, I) \pr(Q \given \lecs, I)
\end{split}
\end{align}
where the distributions for $\cbar^2$ 
and $Q$ are explained in Sec.~\ref{sec:estimatinQ}.
We obtain the left-hand side of Eq.~\eqref{eq:master_posterior} using MCMC sampling. It is then simple to look at projections of these samples for the set of variables one is interested in. This is equivalent to integrating out (or marginalizing over) the other parameters.
This allows us to compute a posterior for $\cD$ and $\cE$ without assuming that the $\NN{}$ LECs or the truncation error parameters are known in advance.
The prior information $I$ that determines the factors in Eq.~\eqref{eq:master_posterior} other than the likelihood, \ie, the prior pdfs, will be discussed in the next subsections.
 
\subsection{Priors for the NN and 3N LECs}
\label{sec:LECpriors}

The prior information $I$ includes $\NN{}$ scattering data, specific values of the $\piN{}$ LECs, and naturalness for $\cD$ and $\cE$. 
The prior on $\lecs \equiv \{\cD, \cE, \lecs_\NN \}$ then factorizes into a prior on the $\NN{}$ LECs, $\lecs_\NN$, 
and one on the \TNF LECs, $\cD$ and $\cE$:
\begin{align}
    \pr(\lecs \given I) & = \pr(\cD, \cE \given I) \pr(\lecs_\NN \given I) \\
    \pr(\cD, \cE \given I) & = \normal[0, \abar^2] \label{eq:lecsprior}\\
    \pr(\lecs_\NN \given I) & = \normal[\mu_\NN, \Sigma_\NN].
    \label{eq:nn_prior}
\end{align}
The bespoke analysis of \NN\ data described in Sec.~\ref{sec:optimization} produces a Gaussian posterior that is our prior on $\lecs_\NN$ for this \TNF{} analysis. We denote the mean and covariance matrix obtained in Sec.~\ref{sec:optimization} by $\mu_\NN$ and $\Sigma_\NN$. 
We adopt a Gaussian for the \TNF\ LEC prior~\cite{Schindler:2008fh,Wesolowski:2016int}.
Its width is chosen as $\abar = 5$. We have found that this value of $\abar$ is sufficiently large that it does not meaningfully impact our full results~\cite{Wesolowski:2018lzj}. 

A fit in which the $\piN{}$ LECs were also constrained by $\NN{}$ and few-body data could be described using the same formalism, by expanding the vector $\lecs$ so that it includes the three $\piN{}$ LECs that appear in the $\NN{}$ potential.

\subsection{Priors for the truncation-error parameters}

\label{sec:estimatinQ}

We now develop the pdf $\pr(\cbar^2, Q \given \lecs, I)$ that enters in Eq.~\eqref{eq:master_posterior}.
This distribution is obtained from two distinct sources of information: (1) the order-by-order pattern of terms in the EFT expansion---%
knowledge of which is implicit in the conditioning on $\lecs,I$%
---and (2) the prior information on $\cbar^2, Q$.
If there were no reliable convergence pattern, or if we happened to be fitting an EFT at leading order, then this pdf would simply reduce to the prior on $\cbar^2, Q$.
For a detailed explanation of this approach, see~\cite{Melendez:2019izc}, the appendices in particular.

Let us begin with a description of how the convergence pattern for $\genobsthset$ enters our analysis.
Again, $\genobsthset$ consists of the mass and radius of ${}^4$He, and the mass and $\beta$-decay rate of ${}^3$H.
For the LO-NLO correction, we have in principle the results in Table~\ref{tab:orderbyorder}.
However, the shift from LO to NLO in the nuclear binding energies is large, being 100\% of the LO value in many cases.
This is because these states are weakly bound, \ie, $\langle T \rangle$ and $\langle V^{(0)} \rangle$  are each much larger in size than the energy, $E$, of the ${}^3$H or ${}^4$He eigenstate.
Therefore while $\langle V^{(2)}\rangle \ll \langle V^{(0)}\rangle$, in accord with \chiEFT\ counting, $\langle V^{(2)} \rangle$ can still be a sizable fraction of the leading-order eigenenergy, $E^{(0)}$.
However,  $E^{(3)}-E^{(2)} \approx \langle V^{(3)} \rangle$, therefore the NNLO shift of the eigenenergy should provide information on the expansion parameter.
Since the radii of weakly-bound states are correlated with the distance they lie from the nearest particle-removal threshold~\cite{Chen:1999tn,Braaten:2002er,Platter:2005sj,Forssen:2017wei} the (large relative) shift in that observable at NLO also does not give straightforward information on the convergence of the \chiEFT\ expansion.

Meanwhile, the convergence pattern for $fT_{1/2}$ of ${}^3$H is irregular: $fT_{1/2}$ receives zero correction at relative order $Q$, while the one-body-operator corrections at $O(Q^2)$ produce a $< 1$\% effect. But a significant alteration to the LO result comes when two-body axial currents appear at $O(Q^3)$~\cite{Baroni:2016aa}. 
The statistical model employed here assumes a regular order-by-order convergence of observables.
More work, \eg, a simultaneous treatment of corrections to the tri-nucleon wave function and the Gamow-Teller operator, is needed to understand why $fT_{1/2}$ does not have such a convergence pattern. But, in the meantime, the order-by-order behavior of $fT_{1/2}$ is not consistent with our statistical model, so we
do not use it to develop the pdf $\pr(\cbar^2, Q \given \lecs, I)$.
The information on the \chiEFT\ convergence pattern that goes into that pdf is from a subset of the order-by-order \chiEFT\ predictions: the NLO-NNLO corrections for $E({}^4\text{He})$, $r({}^4\text{He})$ and $E({}^3\text{H})$.

Now the NLO-NNLO corrections for these observables vary with $\lecs$. The NLO observable calculations are performed using the NLO optimum for the \NN{} LECs (see Sec.~\ref{sec:optimization}).
But the NNLO observables depend on $\lecs$, so we should infer the NNLO LECs and the truncation error parameters simultaneously during the NNLO fit.

We follow~\cite{Melendez:2019izc} and use a scaled inverse chi squared distribution for 
the prior $\pr(\cbar^2 \given I) \sim \invchisq[\nu_0, \tau_0^2]$. This pdf depends on two hyperparameters $\nu_0$ and $\tau_0^2$ that are chosen at the beginning of the analysis. (The prior pdfs we take for $\cbar^2$ and $Q$ are shown as the blue lines in Fig.~\ref{fig:hyperparameter_distributions}.)
Because this is a conjugate prior, the posterior distribution is obtained analytically as
\begin{align}
   \pr(\cbar^2 \given Q, \lecs, I) & \sim \invchisq[\nu, \tau^2(\lecs,Q)] . \label{eq:cbarsqprior}
\end{align}
The updating formulae for the hyperparameters are~\cite{Melendez:2019izc}:
\begin{align}
    \nu & = \nu_0 + N_{\text{obs}} n_c \label{eq:nu}\\
    \nu \tau^2(\lecs, Q) & = \nu_0 \tau_0^2 + \sum_{n,i} c_{n,i}^2(\lecs, Q)\; \label{eq:tausq},
\end{align}
where $i$ indexes the $N_{\text{obs}}$ observables, $n$ indexes the $n_c$ lower order coefficients used to estimate the truncation error, and the observable coefficients are given by
\begin{align}
    c_{n,i}(\lecs, Q) = \frac{\genobs_n^{(i)}(\lecs_{(n)}) - \genobs_{n-1}^{(i)}(\lecs_{(n-1)})}{\genobsref Q^n}.
\end{align}
The notation $\lecs_{(n)}$ describes the LECs found at the $n$th order EFT fit.
For NNLO, these are the $\lecs$ that are varied in the fit, whereas for NLO these are fixed at the optimum from the NLO fit.

With these updated hyperparameters in hand we can then obtain the unnormalized $Q$ posterior:
\begin{align} \label{eq:Q_posterior_unnorm}
    \pr(Q \given \lecs, I) \propto \frac{\pr(Q \given I)}{\tau^\nu \prod_{n} Q^{N_{\text{obs}}n}}.
\end{align}
The fact that Eq.~\eqref{eq:Q_posterior_unnorm} is unnormalized would not usually be a problem for estimating $Q$.
But the normalization factor depends on $\lecs$ because the NLO-NNLO correction depends on $\lecs$.
The set of LECs $\lecs$ is the quantity we are trying to estimate in Eq.~\eqref{eq:master_posterior}, so we must be careful to include this factor.
We quickly normalize Eq.~\eqref{eq:Q_posterior_unnorm} at each MCMC step by precomputing \verifyvalue{70} Gaussian quadrature locations $Q_i$ and weights.
Additional speedup is realized by parallelizing the calls to Eq.~\eqref{eq:Q_posterior_unnorm} across the Gaussian points $Q_i$.

The last ingredient we need is then the prior $\pr(Q \given I)$ that goes into the convergence-pattern analysis. To formulate that we note:
\begin{itemize}
    \item $Q$ is restricted to the range $(0, 1)$;
    \item for properties of low-energy bound states we expect \chiEFT\ to converge with $Q$ less than 1/2~\cite{Binder:2018pgl,Epelbaum:2019kcf,Maris:2020qne}.
\end{itemize}
We encode this as a weakly informative Beta distribution $\text{B}(Q \given \verifyvalue{a=3, b=5})$, which provides a slight bias towards $Q < 0.5$ and has support only for $Q \in (0, 1)$.

With $\pr(Q \given \lecs, I)$ and $\pr(\cbar^2 \given Q, \lecs, I)$ in hand the desired pdf is straightforwardly obtained via the product rule for conditional probabilities:
\begin{equation}
    \pr(\cbar^2,Q \given \lecs,I)=\pr(\cbar^2 \given Q, \lecs, I) \pr(Q \given \lecs,I).
\end{equation}

\subsection{Prior for NN LECs from NN scattering data}
\label{sec:optimization}
We acquire values for the \NN{} sector LECs $\lecs_{\NN}$ at LO, NLO,
and NNLO by performing a new fit to $np$ and $pp$ scattering data in
the $0 < E \leq 290$\,MeV range gathered from the Granada 2013
database~\cite{Perez:2013-08, Perez:2013-12}. As the LEC
$\widetilde{C}_{1S0}^{nn}$ is unconstrained by the scattering data we
also include the empirical $^1$S$_0$ $nn$ scattering length $a_{nn}^N
= -18.95\pm0.40$\,fm and effective range $r_{nn}^N = 2.75\pm0.11$\,fm~\cite{Machleidt:2011zz}.
The optimization procedure maximizes the
likelihood function defined in Eq.~\eqref{eq:likelihood}. Fully
specifying the likelihood requires us to pick values for $\bar{c}$ and
the \NN{} observable expansion parameter $Q_\NN{}(p)$; these are set
to $\bar{c} = 1$ and $Q_\NN{}(p) = \text{max}(m_\pi, p)/\Lambda_b$
where $m_\pi$ is the pion mass, $p$ is the center-of-mass momentum of
the \NN{} system, and $\Lambda_b = 600$\,MeV. A set of reference values $y_\text{ref}$ are also required, for which we use the experimental values.

Three \piN{} LECs ($c_1$, $c_3$, and $c_4$) enter at NNLO\@.
While these LECs could in principle be determined in the same way as the \NN{} LECs, a more precise determination is possible by performing a Roy-Steiner analysis of \piN{} scattering data~\cite{Hoferichter:2015,Hoferichter:2016}. Here we keep the $c_i$'s fixed to the central values from a Roy-Steiner analysis performed by Siemens \etal.~\cite{Siemens:2017} as we focus on the uncertainties from the \NN{} sector. The covariance matrix for the \piN{} LECs provided in Ref.~\cite{Siemens:2017} could straightforwardly be included as prior information in Eq.~\eqref{eq:nn_prior}, provided the cross-correlation between the \piN{} and \NN{} LECs were known.
The fixed values of the $c_i$'s are shown in Table~\ref{tab:nn_lec_values} in Appendix~\ref{sec:optimized_parameter_values}.

The result of an optimization can (and usually does) depend strongly on the choice of starting point $\lecs_0$. A previously found optimum---produced by performing a fit to phase shifts using \textsc{POUND}er\textsc{S}~\cite{wild2014, munson2012} optimization---serves as a basis for choosing a starting point. We choose $\lecs_0$ by randomly perturbing a subset of the previously found parameter values.

With the setup complete we run the optimization using the first-order Levenberg-Marquardt algorithm. This is repeated 600 times using different starting points. One or more candidate optima are chosen and used as starting points to the second-order Newton-CG method, which increases the precision of the found optimum. The final optimum is then chosen as the set of LECs $\lecs^*_\NN{}$ which produces the maximum likelihood value. The resulting values for the LECs agree well with findings from similarly regulated potentials~\cite{Epelbaum:2004fk,Machleidt:2011zz} and  are shown in Table~\ref{tab:nn_lec_values}.

To estimate the covariance matrix $\Sigma_\NN$ of the \NN{} LECs
$\lecs_\NN{}$ we follow the method detailed by Carlsson \etal.\ in
Sec.~IIG of Ref.~\cite{Carlsson:2015vda}. The resulting Gaussian
pdf~\eqref{eq:nn_prior} is shown in green in Fig.~\ref{fig:ec_training_points}. The Hessian needed to compute the covariance matrix, and the first- and second-order derivatives used by the optimization algorithms, are computed to machine precision using automatic differentiation~\cite{charpentier2009}.

\begin{table*}
    \begin{ruledtabular}
    \begin{tabular}{ldddcdc}
         &  \multicolumn{1}{c}{LO} & \multicolumn{1}{c}{NLO} & \multicolumn{1}{c}{$\langle$NNLO$\rangle_{\text{ppd}}$} & \multicolumn{1}{c}{Experiment} &
         \multicolumn{1}{c}{Adopted uncertainty} &
         \multicolumn{1}{c}{$\Delta$NNLO$_\mathrm{ppd}$}\\
    \colrule  \bigstrut[t]       
    $E ({}^3{\rm H})$ [MeV] & -5.65 & -8.38 & -8.52 & $-$8.482~\cite{Purcell:2010hka} & 0.015 & $[-8.613, -8.453]$ \\
    $E({}^4{\rm He})$ [MeV]     & -24.08 & -30.21 & -28.19 & $-$28.296~\cite{Tilley:1992zz} & 0.005 & $[-28.670, -27.853]$  \\
    $r({}^4{\rm He})$ [fm]     & 1.27 & 1.33 & 1.45 & 1.4552(62)~\cite{angeli:2013epw} & 0.0062 & $[1.4414, 1.4634]$  \\
      $fT_{1/2}$ [s]     &  & & 1127.3 & 1129.6(3.0)~\cite{akulov2005} &
        3.0 &      
      $[1109.1, 1150.9]$
    \end{tabular}
    \end{ruledtabular}
    \caption{Results at LO, NLO, and NNLO for the observables used in various combinations to form our likelihood: the binding energies of the ${}^3$H and ${}^4$He states, the rms (point-proton) radius of ${}^4$He and the $\beta$-decay comparative half-life of ${}^3$H. Experimental data are from~\cite{Tilley:1992zz,Purcell:2010hka, akulov2005, angeli:2013epw}. Non-negligible uncertainties in the last digits are then given in brackets. Adopted uncertainties are the larger of those and uncertainties from the calculational method used to solve the Schr\"odinger equation. Note that corrections have been applied to experimental data to obtain the third and fourth ``observables,'' as described in the text.
    The $\langle$NNLO$\rangle_\text{ppd}$ results were obtained by averaging over the Bayesian posterior predictive distribution (ppd) for the EFT predictions; see Eq.~\eqref{eq:posterior_predictive_distribution} and Fig.~\ref{fig:posterior_predictive}.
    The \verifyvalue{68\%} highest posterior density (HPD) credible regions of the NNLO predictions are shown in the $\Delta$NNLO$_\text{ppd}$ column.
    }  
    \label{tab:orderbyorder}    %
  \end{table*}

 \section{Few-nucleon-physics implementation%
   \label{sec:fewbodydetails}}
 The likelihood in Eq.~\eqref{eq:likelihood} is centered at the model predictions ${\bf \genobsth}$ for few-nucleon ($A=3,4$)  observables. To make those predictions we apply the No-Core Shell Model (NCSM)~\cite{Navratil:1999pw} in a  relative-coordinate harmonic-oscillator (HO) basis and solve the few-nucleon Schr\"odinger equation with two- and three-nucleon interactions employing the isoscalar approximation as presented in Ref.~\cite{kamuntavicius1999}. The model-space dimension is determined from the truncation in total number of HO excitations $N_{\rm max}$.
The eigenenergy of the resulting Hamiltonian matrix is a variational estimate of the total binding energy while the eigenfunctions can be used to obtain other observables.

We obtain converged ground-state observables using $\hbar \omega=36$\,MeV and $N_{\rm max}=40(18)$ for $A=3(4)$ since we employ a rather soft chiral interaction at NNLO\@.
Specifically we use a nonlocal momentum-space regulator function as in Eqs.~(5) and (6) of Ref.~\cite{Carlsson:2015vda} with cutoff $\Lambda=450$\,MeV and $n=3$. For $^4$He we obtain ground-state energies and point-proton radii that are converged within $\lesssim5$\,keV and $\lesssim 0.002$\,fm compared to larger-basis calculations.

\subsection{Few-nucleon observables of interest} 

The first two observables we consider are the binding energies of ${}^3$H and ${}^4$He. Determining these from precisely known masses yields errors on the binding energies of a few eV or less. This is negligible compared to errors from the method used to calculate the bound states. Therefore in Table~\ref{tab:orderbyorder} we take ``adopted errors'' for these two observables of 15\,keV (width of the 68\% credibility interval given the 20\,keV accuracy of the isoscalar approximation for the ${}^3$H binding energy quoted in Ref.~\cite{kamuntavicius1999}) and 5\,keV (NCSM basis truncation) respectively. Ultimately, both of these are dwarfed by the truncation error.

We also compute the point-proton radius, here denoted $r$, for $^3$H and $^4$He 
and relate it
to the measured charge radius via~\cite{Friar:1975pp} 
\begin{equation}
   r^2  = r_\mathrm{ch}^2 - r_p^2 - \frac{N}{Z} r_n^2 - r_\mathrm{DF}^2 - \Delta r^2,
\end{equation}
where $r_p^2$ ($r_n^2$) is the proton (neutron) mean-squared charge radius, $Z$ ($N$) is the proton (neutron) number, and $r_\mathrm{DF}^2 = 3 \hbar^2 / ( 4 M_p^2) \approx 0.033$~fm$^2$ is the Darwin-Foldy correction~\cite{jentschura2011}. 
There are two-body-current and further relativistic corrections to $r({}^4{\rm He})$ at orders beyond NNLO in \chiEFT, but these are accounted for by the truncation uncertainties in our likelihood, 
so we set $\Delta r^2 = 0$.
We use $r_p = 0.8783(86)$~fm and $r_n^2 = -0.1149(27)$~fm$^2$~\cite{angeli:2013epw}.
We do not use the ${}^3$He binding energy or point-proton radius for inference because they are highly correlated with the corresponding ${}^3$H observables: 100\% correlated in the limit of isospin-symmetric interactions.

Furthermore, we use the triton half-life to provide a constraint on the nuclear force from an electroweak observable. We follow the approach by \citet{Gazit:2009PRL} and compute the triton half-life from the reduced matrix element for $E_{1}^{A}$, the $J = 1$ electric multipole of the axial-vector current
\begin{align}
    \left\langle E_1^A\right\rangle \equiv \big|\!\left\langle
    {}^3\text{He}\middle\|E_1^A\middle\| {}^3\text{H}\right\rangle\!\big|.
\end{align}
Due to the \chiEFT\ link between electroweak currents in nuclei and
the strong interaction
dynamics~\cite{Park:2002yp,Gardestig:2006hj,Gazit:2009PRL}, this
matrix element has a term proportional to $\cD$---the LEC that also
determines the strength of the one-pion-exchange plus contact
interaction diagram of the \TNF. (Note, though, that Krebs has recently pointed out
that this connection is broken at subleading order by commonly used regulation procedures~\cite{Krebs:2020pii}.)
The experimental value for the
comparative half-life, $fT_{1/2} = 1129.6 \pm 3$\,s~\cite{akulov2005},%
\footnote{
Reference~\cite{Baroni:2016aa} uses the value $fT_{1/2} = 1134.6 \pm 3$\,s, obtained from Simpson's tritium $\beta$-decay measurement~\cite{Simpson:1987zz}. The difference between the two $fT_{1/2}$ numbers is larger than the stated error in either. Here we select the Akulov-Mamyrin result, but the tools we have developed and provide make it straightforward to re-do the analysis using either the Simpson value or a compromise $fT_{1/2}$ with an error  inflated so it is large enough to accommodate both results.
}
leads to an empirical value for $\langle E_1^A\rangle = 0.6848\pm 0.0011$~\cite{Gazit:2009PRL} via the relation
\begin{align}
  fT_{1/2} = \frac{K/G_V^2}{(1-\delta_c) + 3 \pi (f_A/f_V)\langle E_1^A \rangle^2},
\end{align}
with $K/G_V^2 = 6146.6 \pm 0.6$\,s, $f_A/f_V = 1.00529$, and the isospin-breaking correction  $\delta_c = 0.13\%$.

%
Results at LO, NLO, and NNLO for these four $A=3,4$ observables,
together with the experimental numbers, are given in Table~\ref{tab:orderbyorder}.
The NNLO results in this table are the mean values obtained from the posterior predictive distribution; see Eq.~\eqref{eq:posterior_predictive_distribution} and Fig.~\ref{fig:posterior_predictive}.

\subsection{Efficient emulators for few-nucleon observables}
\label{sec:emulator}

Although we are studying $A=3,4$ systems using soft interactions, the matrix representations of the  NCSM Hamiltonians for the few-nucleon states that we analyze still reach dimensions of approximately $10^4 \times 10^4$. With a Lanczos algorithm it takes about one minute, using a single CPU, to obtain the energies and corresponding wavefunctions for the systems of interest. 
It takes a few hours of computation on a single node to fully sample the posterior pdf $\pr(\cD, \cE \given \genobsexpset, I)$.

\begin{figure}[t]
    \centering
    \includegraphics{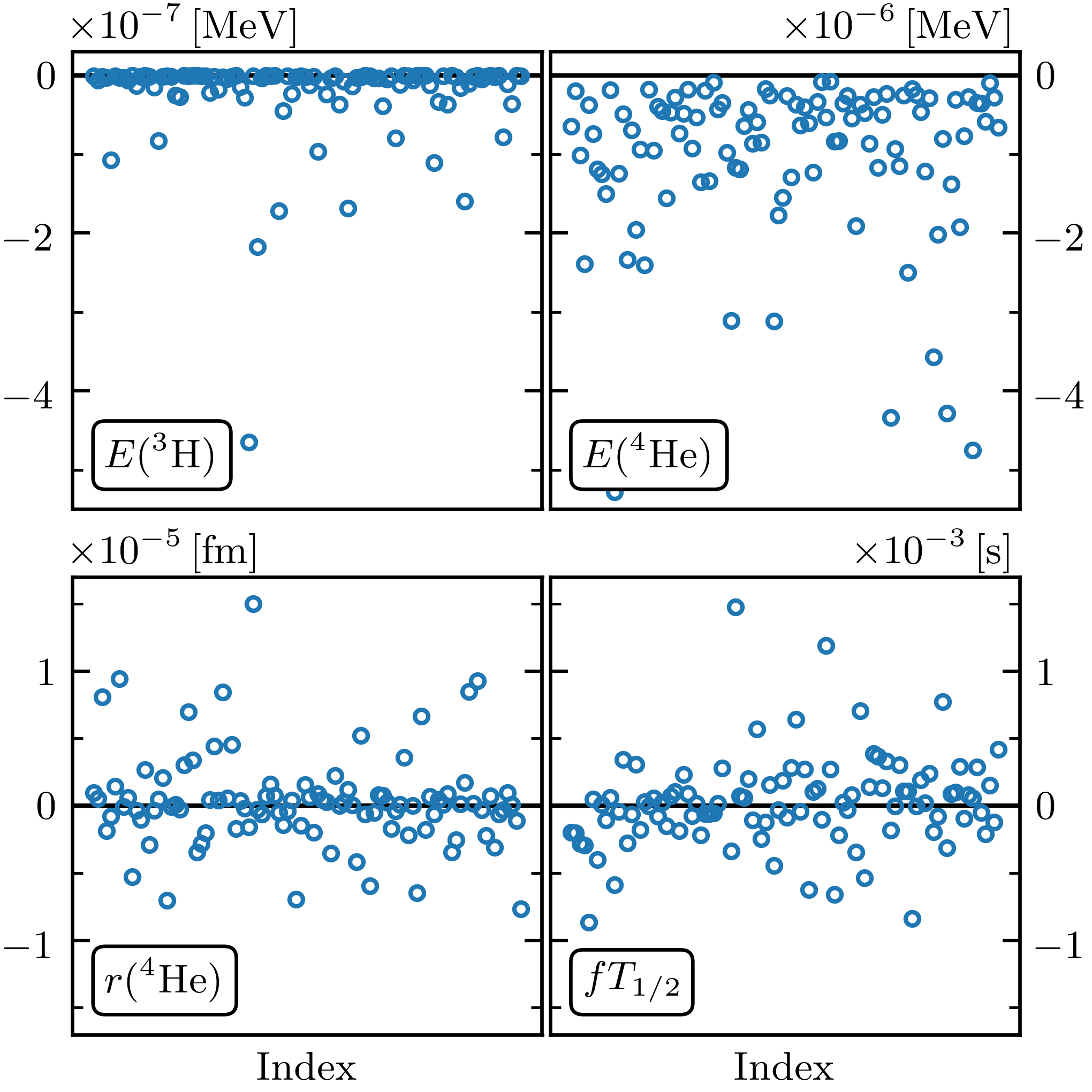}
    \caption{Differences between evaluations using the Schr\"odinger-equation solution and those using the \EC{} emulator for the four observables of interest at \verifyvalue{100} validation points. These differences  are several orders of magnitude smaller than the adopted errors in Table~\ref{tab:orderbyorder}. Note that the ground state energies from the emulator are guaranteed to be an upper bound on the exact energies, but the other observables have no such constraint.
    }
    \label{fig:validation_residuals}
\end{figure}

To enable more rapid iterations of our exploratory data analysis,
we employ eigenvector continuation (EC)~\cite{Frame:2017fah} to efficiently and accurately emulate~\cite{Konig:2019adq} the
$\lecs$ dependence
of the few-nucleon observables listed in Table~\ref{tab:orderbyorder}.
The high accuracy achieved is demonstrated by the smallness of the differences between the emulator and the NCSM result; see Fig.~\ref{fig:validation_residuals}.
The evaluation of the posterior is dramatically accelerated via the EC emulators such that each parameter sample only takes ${\sim 10}$\,ms on a single-threaded CPU with a corresponding speed-up for sampling the relevant parameter space of LECs.
In addition, the construction of a set of model-specific emulators allows others to easily reproduce, and modify, our statistical analysis.\footnote{The NCSM emulators and statistical models can be obtained or created via our open-source python package \texttt{fit3bf}~\cite{fit3bf}.}

The \EC{} approach to emulation is described in Ref.~\cite{Konig:2019adq}; 
to be self-contained, we briefly outline the method here.
Consider one quantum system that we want to emulate, such as the triton.
The $A$-nucleon Schr\"odinger equation can be written as
\begin{equation}
  H(\lecs) \ket{\psi(\lecs)} = E(\lecs) \ket{\psi(\lecs)},
  \label{eq:Ancsm}
\end{equation}
where $\ket{\psi(\lecs)}$ and $E(\lecs)$ denote
the ground-state and its energy, and the implicit
$\lecs$-dependence has been brought forward.
We then diagonalize the Hamiltonian $H(\lecs)$ for \NEC{} different values of $\lecs$, and collect the \NEC{} ground-state wave functions $\ket{\psi_i}$ into a matrix $\ritzbasis$,
\begin{align}
    \ritzbasis \equiv
    \begin{pmatrix}
        \kern.2em\vline\kern.2em & \kern.2em\vline\kern.2em &  & \kern.2em\vline\kern.2em \\
        \ket{\psi_1} & \ket{\psi_2} &  \cdots & \ket{\psi_{\NEC}} \\
        \kern.2em\vline\kern.2em & \kern.2em\vline\kern.2em & & \kern.2em\vline\kern.2em
    \end{pmatrix},
\end{align}
which \emph{does not} depend on $\lecs$.
Then we project the Hamiltonian to a subspace spanned by the \NEC{} wave functions via 
\begin{equation} \label{eq:subspace_proj}
   \subspace{H}(\lecs) = \ritzbasis^\dagger H(\lecs) \ritzbasis .
\end{equation}
Because the chiral Hamiltonians $H$ that we use depend linearly on $\lecs$, this projection can be performed once for each term and stored to quickly construct $\subspace{H}(\lecs)$.

To construct an emulator for $\ket{\psi(\lecs)}$ and $E(\lecs)$, we solve the $\NEC \times \NEC$ generalized eigenvalue equation
\begin{align} \label{eq:ec_generalized_problem}
  \subspace{H}(\lecs) \coeff(\lecs) = \subspace{E}(\lecs) \normmat \coeff(\lecs),
\end{align}
where 
$\normmat = \ritzbasis^\dagger \ritzbasis$ is the norm matrix with elements
$\normmat_{ij} = \braket{\psi(\lecs_i)}{\psi(\lecs_j)}$.
The generalized eigenvalue $\subspace{E}$ is an approximation to the true eigenenergy.
The length-\NEC{} vector of coefficients $\coeff(\lecs)$ found by solving Eq.~\eqref{eq:ec_generalized_problem} could then be used to reconstruct the approximate wave functions via $\ket{\psi(\lecs)} \approx \ritzbasis\coeff(\lecs)$, but these are not needed in practice. Instead, to evaluate expectation values of observables $\hat O$ other than nuclear spectra, one computes
\begin{align} \label{eq:ec_expectation_emulator}
    \ev*{\hat O(\lecs)} & = \mel{\psi(\lecs)}{\hat O(\lecs)}{\psi(\lecs)} \notag\\
    & \approx \coeff(\lecs)^\dagger [\ritzbasis^\dagger \hat O(\lecs) \ritzbasis] \coeff(\lecs).
\end{align}
If $\hat O(\lecs)$ is linear in $\lecs$ then the terms in $\ritzbasis^\dagger \hat O(\lecs) \ritzbasis$ can again be computed once and stored prior to sampling.
For the $\beta$-decay transition, we generalize Eq.~\eqref{eq:ec_expectation_emulator} to the case where the right and left $\ritzbasis\beta(\lecs)$ come from the initial- and final-state emulators, respectively.
This is the first application of EC emulation to a nuclear transition.

\begin{figure}[tb]
    \centering
    \includegraphics{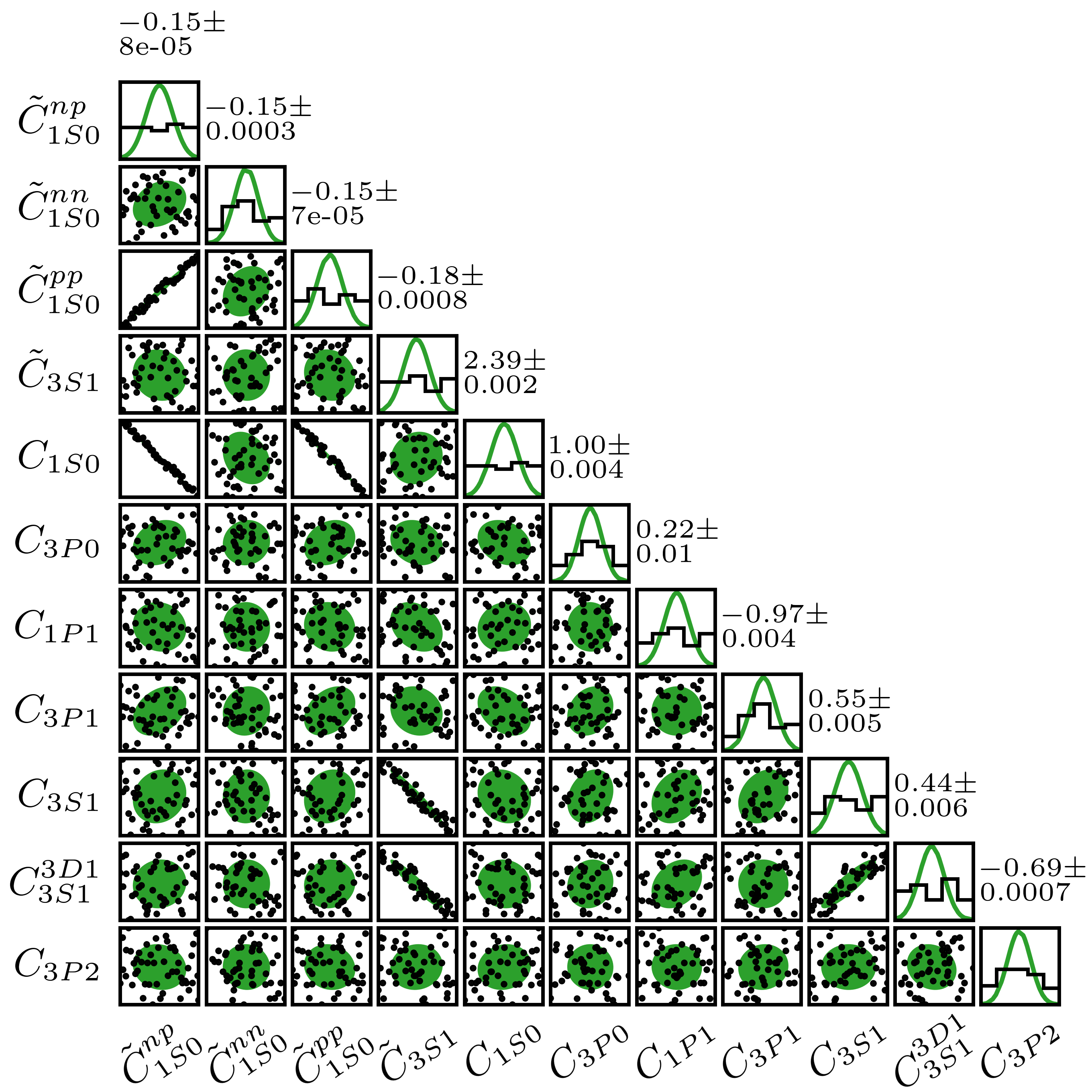}
    \caption{The \EC{} training points in the space of \NN{} LECs compared to the input priors for our few-body analysis. The green curves represent the marginal prior distributions, with the ellipses denoting 95\% credibility regions.
    Each dot is a training point; the marginals are approximately uniformly distributed. The values quoted above the diagonal are means and standard deviations from Table~\ref{tab:nn_lec_values}.
    }
    \label{fig:ec_training_points}
\end{figure}

It was shown in
Ref.~\cite{Konig:2019adq} that $\subspace{E}$ approximates $E$
extremely well even with a small number of training vectors.
Although the Hamiltonian eigenvector originally resides in a Hilbert space of very large dimension, the eigenvector trajectory produced by continuous changes of the Hamiltonian matrix can be accurately represented in a space of very low dimension. For this reason we 
can construct fast
and accurate emulators for all observables that we study, including the $\beta$-decay transition (see Fig.~\ref{fig:validation_residuals}). 

As already noted,
to construct a \emph{computationally efficient} \EC{} emulator requires that we can write the
subspace-projected Hamiltonian as a linear combination of the continuous parameters that
we are interested in. 
For example, considering only the $\cD$ and $\cE$ dependence, we can express the
chiral NNLO Hamiltonian as
\begin{equation}
  H(\cD, \cE) = H^{(\rm const)} + \cD V^{(1\pi-{\rm ct})} + \cE V^{(\rm \NNN-ct)} 
  ,
  \label{eq:Hnnlo}
\end{equation}
where we partitioned the Hamiltonian into three pieces:
all contributions that are constant with respect to variation of $\cD$
and $\cE$ ($H^{(\rm const)}$), the one-pion-exchange plus contact ($1\pi$--ct) interaction between three nucleons, and the  pure
three-nucleon contact (\NNN--ct). 
Having obtained  \NEC{} linearly independent training vectors
$\ket{\psi_i}$ for each state of interest, we construct each subspace-projected matrix [denoted with tildes as in \eqref{eq:subspace_proj}] in
\begin{equation}
  \subspace{H}(\cD, \cE) = \subspace{H}^{(\rm const)}+ \cD\subspace{V}^{(1\pi-{\rm ct})} +\cE\subspace{V}^{(\rm \NNN-ct)}
\end{equation}
only once prior to sampling, which greatly speeds up the subsequent matrix algebra.
Equipped with the subspace basis, we can also project the
operators for the point-proton radius of $^{4}$He and triton
$\beta$-decay.

When sampling over both \NN{} and 3N LECs (for a total of 13 dimensions), we use $\NEC = \verifyvalue{50}$ training points.
For the \TNF{} LECs, we simply use a Latin hypercube design in the range $[-5, 5]$.
For the \NN{} LECs, we start with a Latin hypercube design in the range $[-1, 1]$.
We then map each training point $p_i$ to the plausible range of LECs according to $\mu_{\NN} + \Sigma_{\NN}^{1/2}p_i$
with $\mu_{\NN}$ and $\Sigma_{\NN}$ the mean and covariance determined in Sec.~\ref{sec:optimization}.
The resulting set of points form our training LECs, and are displayed in Fig.~\ref{fig:ec_training_points}.


\section{Bayesian parameter estimation for \texorpdfstring{$\cD$}{cD}, \texorpdfstring{$\cE$}{cE}, \texorpdfstring{$Q$}{Q}, and \texorpdfstring{$\cbar$}{cbar}} \label{sec:results}

\begin{figure}[tb!]
    \centering
    \includegraphics{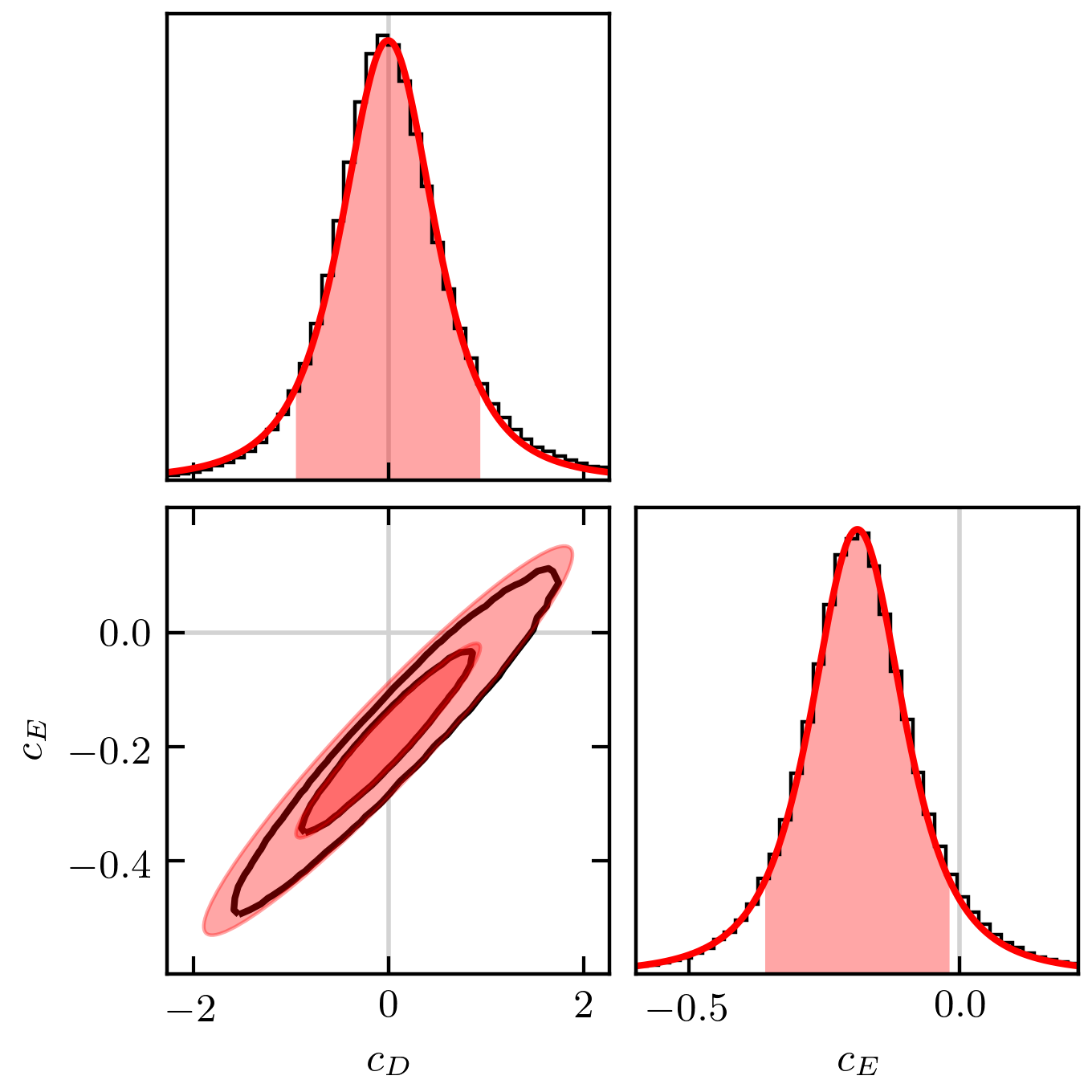}
    \caption{The posterior of $\cD$ and $\cE$ fitting to all four few-body observables and marginalizing over $\cbar^2$, $Q$, and the \NN{} LECs.
    The black histograms and contours correspond to the pure MCMC samples.
    The red curves and ellipses follow from a fit of a multivariate $t$ distribution $t_\nu(m, S)$ as described in the text.
    Filled areas in the marginals denote one standard deviation of the fit distribution, which contains \verifyvalue{$86\%$} of the probability mass, not $68\%$ like a Gaussian.
    Contours represent the one and two standard deviations of the best fit $t_\nu(m, S)$.
    }
    \label{fig:posterior_cd_ce}
\end{figure}

\begin{figure}
    \centering
    \includegraphics[width=\columnwidth]{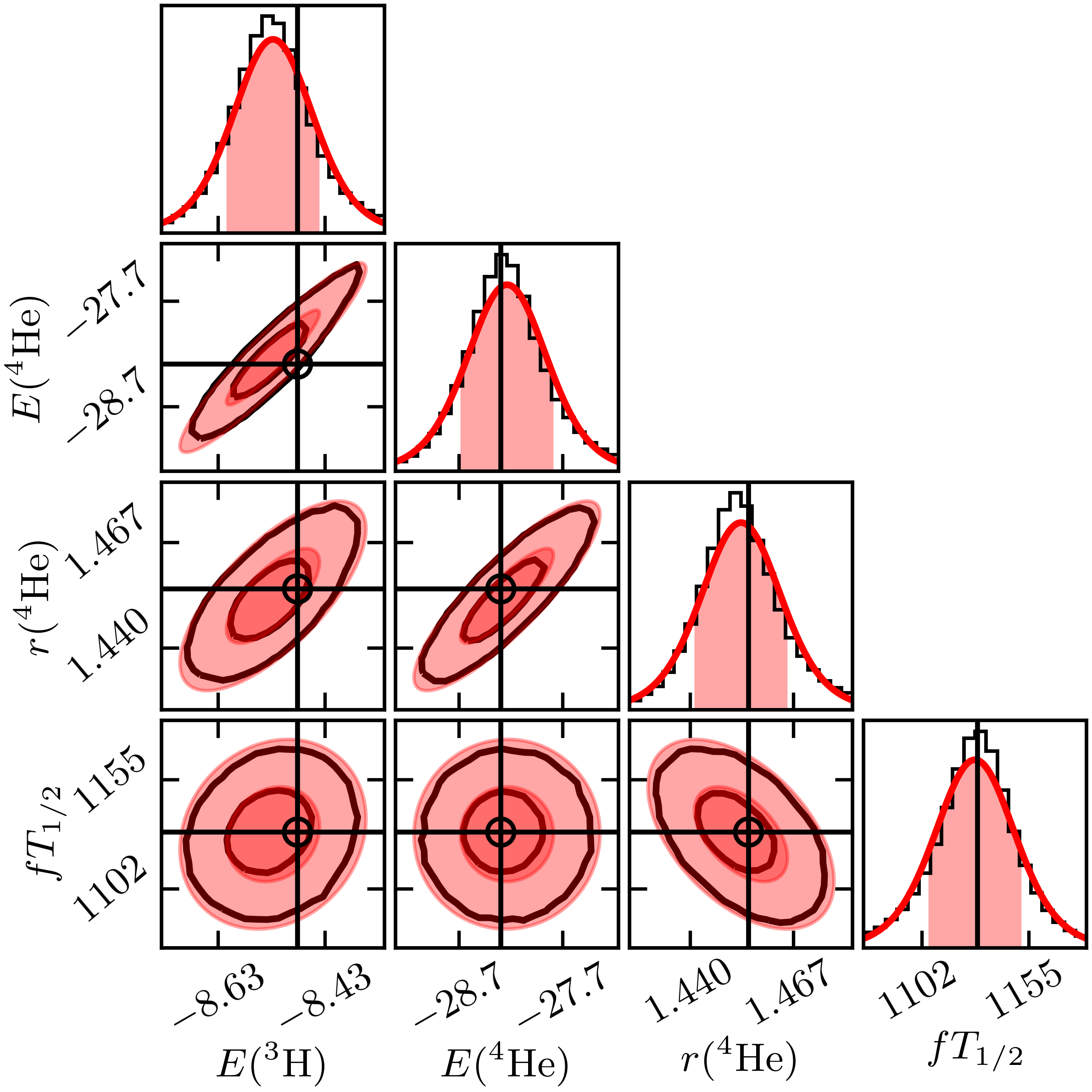}
    \caption{The posterior predictive distribution from sampling over the LECs found in Fig.~\ref{fig:posterior_cd_ce}, with units as in Table~\ref{tab:orderbyorder}.
    The red distributions come from a fit of a multivariate $t$ distribution to the data (see Appendix~\ref{app:to_a_t}).
    The filled regions of the 1d plots represent one standard deviation of the marginal $t$ distributions.
    The filled contours of the joint distributions denote the 1 and 2 standard deviation regions of the multivariate $t$, and the black contours denote the corresponding HPD regions from the samples.
    The markers and black horizontal and vertical lines denote the experimental values.
    }
    \label{fig:posterior_predictive}
\end{figure}

\begin{figure}
    \centering
    \includegraphics{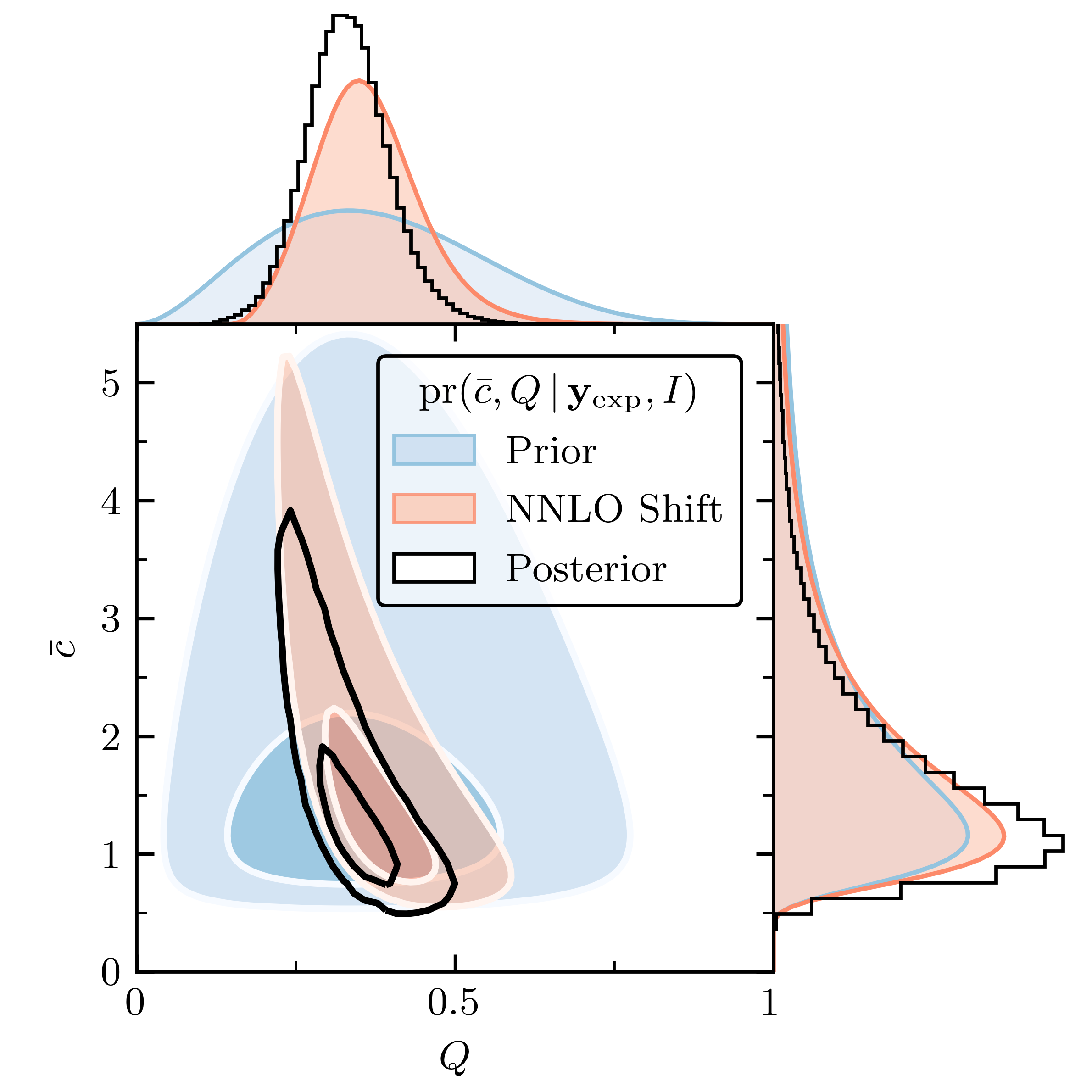}
    \caption{The prior and posterior distributions for $\bar c$ and $Q$.
    The blue line denotes the uncorrelated prior distribution with $\cbar^2 \sim \invchisq[\verifyvalue{\nu_0 = 1.5, \tau_0^2 = 1.5^2}]$ and $Q \sim \text{B}(\verifyvalue{a=3, b=5})$.
    The black posterior is obtained by conditioning on the NLO-NNLO shift at each $\lecs$ value in the sampler. It also folds in information about the size N$^3$LO effects need to have to yield agreement with the data.
    From this we obtain $Q = \verifyvalue{0.33 \pm 0.06}$.
    If we instead updated from the prior to the posterior via the mean value for the shift obtained from the fit, then we would have obtained the red curve.
    }
    \label{fig:hyperparameter_distributions}
\end{figure}

\begin{figure*}
    \centering
    \subfloat[]{
    \includegraphics{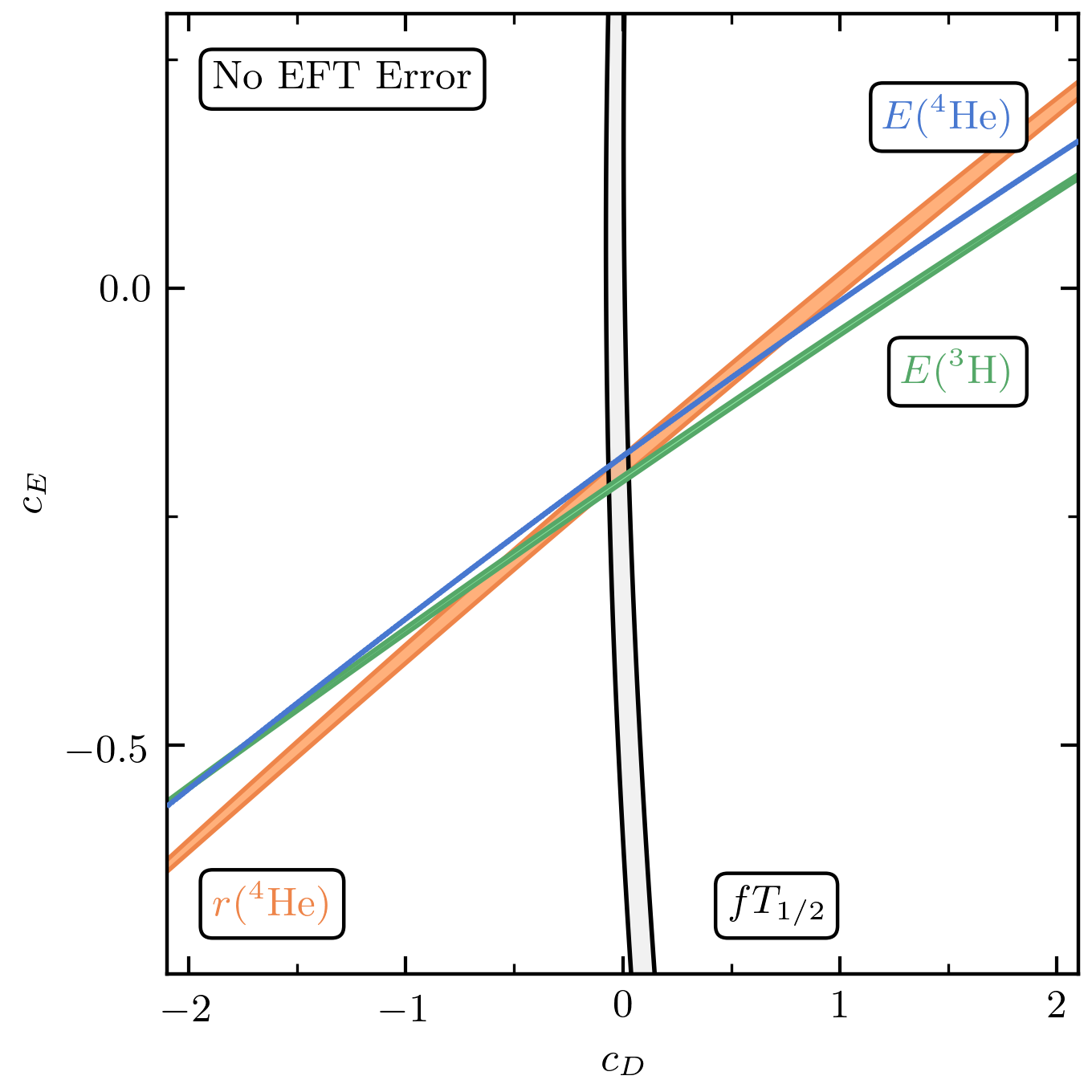}
    \label{fig:single_obs_constraints_no_truncation}
    }
    \subfloat[]{
    \includegraphics{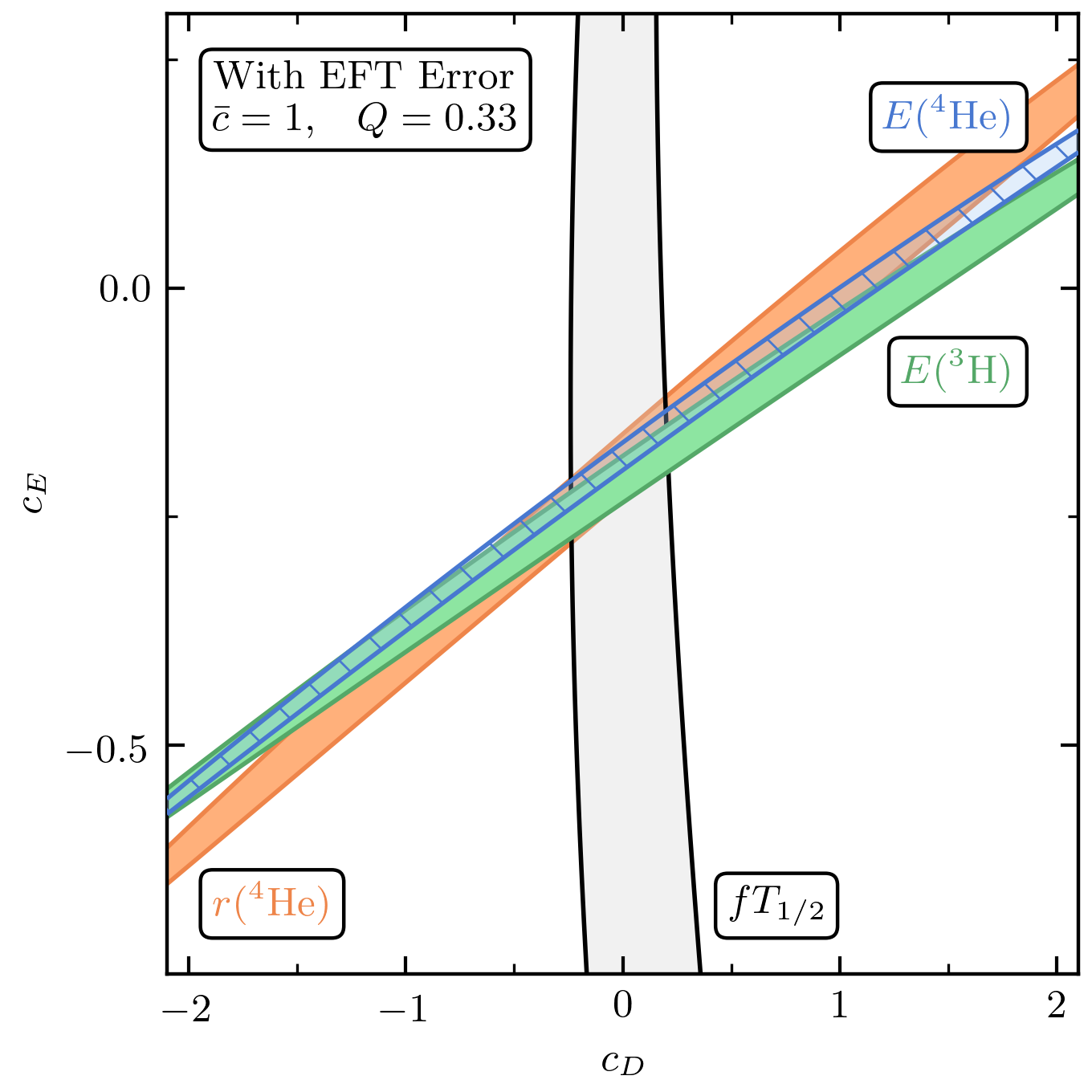}
    \label{fig:single_obs_constraints_with_truncation}
    }
    \caption{
    Constraints on $\cD$ and $\cE$ from single-observable fits both without \protect\subref{fig:single_obs_constraints_no_truncation} and with \protect\subref{fig:single_obs_constraints_with_truncation} EFT truncation errors included.
    The regions are determined by computing 39\% HPD intervals, which would correspond to $1\sigma$ intervals for a 2d Gaussian.
    The \NN{} LECs are fixed to their prior values for these plots, as they make little difference to the overall fit.
    There is no mutual overlap of the truncation-error-free posteriors, which would make a simultaneous fit difficult and unreliable.
    On the other hand, the right-hand panel makes it clear that there is no inconsistency in the theory here, once truncation errors are accounted for.
    (Note that in the right-hand panel $\cbar$ and $Q$ are fixed. If they were allowed to vary as in the full fit, then there is not enough information to accurately constrain these posteriors.)
    All but the $fT_{1/2}$ observable provide essentially identical information about the $\cD$, $\cE$ fit, which makes it a crucial observable to include.
    }
    \label{fig:single_obs_constraints}
\end{figure*}

\begin{figure}
    \centering
    \includegraphics{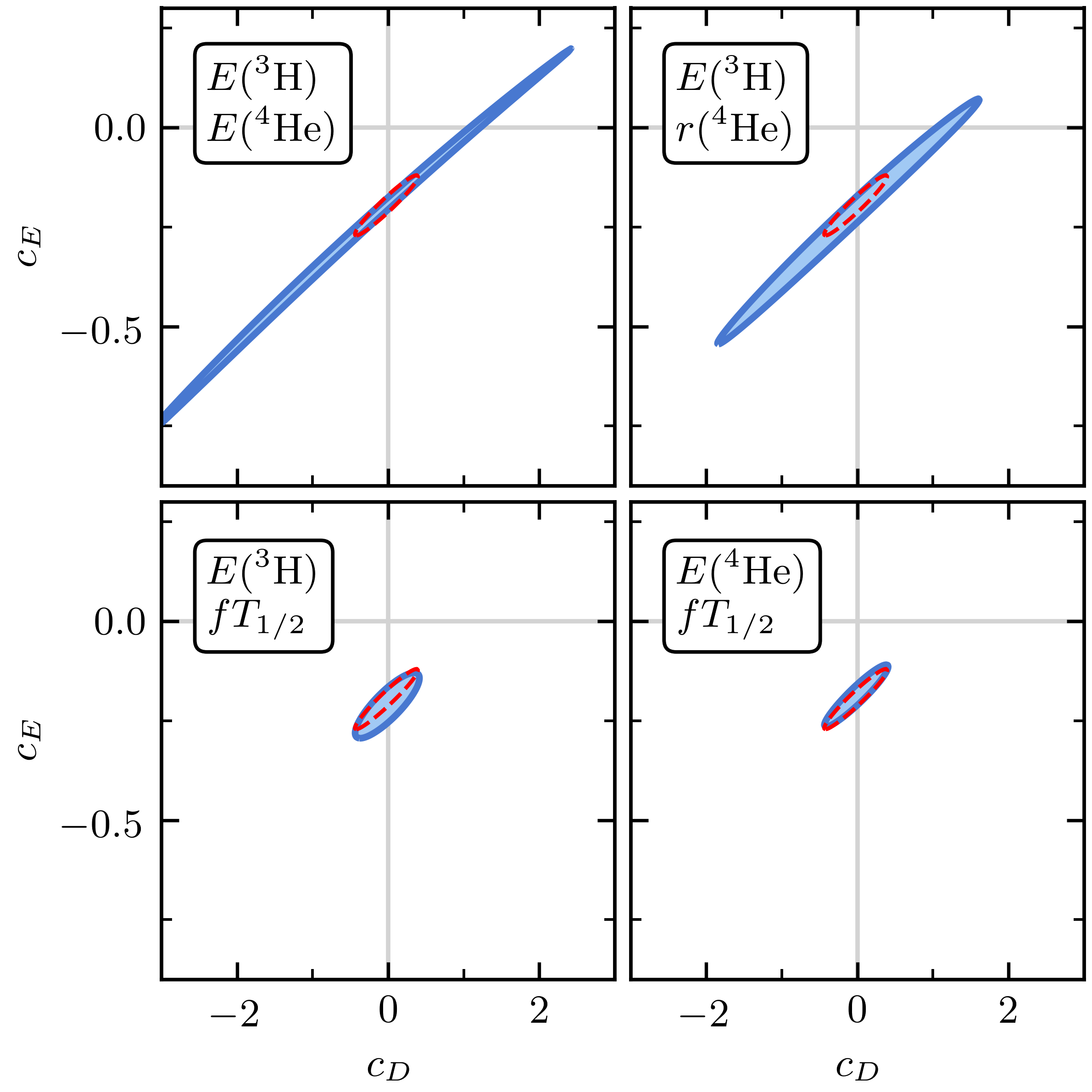}
    \caption{Posteriors found by including only subsets of the
      few-body observables in the likelihood~\eqref{eq:likelihood}
      while holding $\lecs_{\NN}$ and truncation parameters fixed (\verifyvalue{$\cbar=1$, $Q=0.33$}).
    In each panel the fit to all four observables, again with fixed truncation parameters, is represented by the dashed line, and is shown for  comparison.
    We choose to show the 39\% HPD interval, which corresponds closely to $1\sigma$ because these distributions are approximately Gaussian.
    Posteriors conditioned on the triton half-life give particularly well-constrained \TNF parameters, whereas the others are unconstrained along a linear combination of $\cD$ and $\cE$.
    The fit to $E({}^4\text{He})$ and $fT_{1/2}$ produces almost the same posterior as the fit to all four observables.
    Because the truncation parameters $\cbar$ and $Q$ are fixed---for the same reasons as in Fig.~\protect\subref*{fig:single_obs_constraints_with_truncation}---these posteriors appear more constrained than is justified by our true prior knowledge.
    }
    \label{fig:posterior_cd_ce_subsets}
\end{figure}

Figure~\ref{fig:posterior_cd_ce} shows the joint posterior for $\cD$ and $\cE$ as obtained from MCMC sampling of the full posterior \eqref{eq:master_posterior}.
This LEC posterior 
has been marginalized over $\lecs_\NN$ as well as the truncation error parameters $\cbar$ and $Q$.
The evaluation was done using fixed $\abar=\verifyvalue{5}$, although the final posterior is concentrated so close to zero that $\abar$ could be taken to larger values without influencing the results. The data likelihood~\eqref{eq:likelihood} contains the four few-nucleon observables listed in Table~\ref{tab:orderbyorder}.
We sample the posterior using the affine invariant MCMC ensemble sampler \texttt{emcee}~\cite{Foreman:2013} using \verifyvalue{50} walkers with \verifyvalue{50,000} iterations per walker following \verifyvalue{2,000} warmup steps.

The joint distribution in Fig.~\ref{fig:posterior_cd_ce} is best represented by a multivariate $t$ distribution.
The emergence of a $t$ distribution is a generic feature of statistics problems that are linear in the parameters and involve variance estimation---as explained in Appendix~\ref{app:to_a_t}---and the linear correlations seen in Fig.~\ref{fig:posterior_cd_ce_subsets} strongly support that this problem is approximately linear in $\cD$ and $\cE$. None of this is surprising: the \TNF is a perturbative correction in \chiEFT\ and the values of $\cD$ and $\cE$ that turn out to be relevant are small.
(For another recent discussion of the benefits of a perturbative treatment of $\cD$ and $\cE$ see Ref.~\cite{Witala:2021ufh}.)

We fit a parametrized distribution to the $\cD$, $\cE$ samples by maximizing their likelihood given that they are multivariate $t$ distributed $t_\nu(m, S)$.
The best fit is obtained with $\nu \approx \verifyvalue{2.8}$ degrees of freedom, a mean vector
$m = \begin{bmatrix} \verifyvalue{-0.0047} & \verifyvalue{-0.1892} \end{bmatrix}$,
and scale matrix of
\begin{align*}
    S =
    \begin{bmatrix}
        0.250 & 0.043 \\
        0.043 & 0.008
    \end{bmatrix} .
\end{align*}
This yields an accurate description of the one-dimensional $\cD$ and $\cE$ posteriors and of their joint pdf  at one standard deviation. The two standard deviation contour in the two-dimensional LEC pdf is harder to match. 
This distribution has moderately heavy tails---a Gaussian is not a good approximation. 

The parameters $\cD$ and $\cE$ are strongly correlated. The covariance matrix is $\nu S / (\nu - 2)$, corresponding to a correlation coefficient $\rho \approx \verifyvalue{0.96}$.
The strength of this correlation is similar to what was found in Baroni \etal.~\cite{Baroni:2016aa} and Kravvaris \etal.\ in Ref.~\cite{Kravvaris:2020lhp}. In contrast, in Ref.~\cite{Epelbaum:2018ogq} Epelbaum \etal.\ employed SCS potentials and found the triton-binding-energy constraint led to $\cD$ and $\cE$ being anti-correlated. The way that this correlation is connected to the wave function of the three-nucleon system and the short-distance behavior of the \NN{} force is an interesting subject for future study.

The consistency of our parameter estimation can be assessed by studying the model posterior predictive distribution (ppd)
\begin{align} \label{eq:posterior_predictive_distribution}
    \text{ppd} = \{\genobsthset(\lecs) : \lecs \sim \pr(\lecs \given \genobsexpset, I) \}.
\end{align}
The ppd is the set of all predictions computed over likely values of the LECs, \ie, drawing from the posterior pdf for $\lecs$.
Figure~\ref{fig:posterior_predictive} shows the ppd for the target few-nucleon observables, evaluated from the full posterior~\eqref{eq:master_posterior}. In practice, the ppd is evaluated via sampling and we use the MCMC samples of the full posterior for this purpose.
The four target experimental values are within one standard deviation for all of the marginals, while all but one pair of values are within one standard deviation regions for the bivariate joint distributions.
For the ${}^3$H-${}^4$He joint distribution the target is instead within the two standard deviation region.
We reiterate that the probability mass enclosed in these intervals does not correspond to Gaussian intervals due to the heavy tails of the distribution.

Because we simultaneously sample the \TNF LECs and the parameters associated with our model for truncation errors, we also have access to the (joint) posterior for those parameters, $Q$ and $\cbar$. This posterior is shown in Fig.~\ref{fig:hyperparameter_distributions} as the black histogram. It should be compared to the prior distribution represented by the blue curve and described in Sec.~\ref{sec:estimatinQ}.
Both the NLO to NNLO shift in observables and the discrepancies with data of the NNLO \chiEFT\ predictions inform the pdf for $\cbar$ and $Q$.
Together, the constraints yield $Q = 0.33 \pm 0.06$, which is an uncertainty of about 20\%.
An ongoing analysis by the LENPIC collaboration suggests $Q = \{m_\pi\}_{\rm eff} / \Lambda_b$ with $\{m_\pi\}_{\rm eff} \approx 200\,$MeV and $\Lambda_b \approx 600$--650\,MeV, a very similar value for $Q$~\cite{Binder:2018pgl,Epelbaum:2019kcf,Maris:2020qne}. The preferred values of $\cbar$ are of order one: the one-dimensional 68\% Bayesian credible interval is $\cbar \in [0.87,1.44]$. This validates the naturalness assumptions encoded in the truncation-error model. There is a nonlinear correlation between $\cbar$ and $Q$, presumably because the pattern of EFT convergence constrains the combinations $\cbar Q^3$ (from NLO to NNLO) and $\cbar Q^4$ (NNLO uncertainties).

If one could assume reasonable values for $\lecs$ \emph{a priori}, then an alternative approach to this evaluation of the $Q$-$\cbar$ posterior via sampling is to use the mean value of the shift in observables from NLO to NNLO to update the pdf for $\cbar$ and $Q$; see Eqs.~\eqref{eq:cbarsqprior}--\eqref{eq:Q_posterior_unnorm}.
Updating using the mean values from the ppd and the NLO numbers in Table~\ref{tab:orderbyorder} yields the \verifyvalue{red} curves for $Q$ and $\cbar$ in Fig.~\ref{fig:hyperparameter_distributions}. These differ from the sampling results in two ways.
First, in the sampling results the NLO-to-NNLO shift is computed for each sample separately. The value of $c_3$, and hence that of $\tau^2$ and $\cbar^2$, depends on $\lecs$, and so is different for each member of the MC Markov chain.
However, since the ppd of all the observables that inform the convergence pattern is quite narrow, this $\lecs$-dependence is a small effect. 
The samples in Fig.~\ref{fig:hyperparameter_distributions} also account for the requirement that the sizes of the NNLO errors are statistically consistent. The combination $\cbar Q^4$ determines the variance of our NNLO  pdfs. Incorporating NNLO variance estimation in our $\cbar$-$Q$ estimate brings the central value of $Q$ down slightly compared to what is obtained if only the NLO-to-NNLO shift in observables is considered.

If truncation errors are not included in the analysis then the individual constraints from all four observables disagree by several $\sigma$, see Fig.~\subref*{fig:single_obs_constraints_no_truncation}, where \NN{} LECs are also held fixed.\footnote{
In the absence of a prior these posteriors extend very far in both directions, since the problem is approximately linear and each band represents the constraint on two parameters from one datum. But the prior on $\lecs$ [Eq.~\eqref{eq:lecsprior}] regulates these one-dimensional structures once values of $\cD$ and $\cE$ $\approx \abar$ are reached.
}
Consequently, obtaining a posterior with $\covarth=0$ becomes both difficult and unreliable.
In particular, 
the errors adopted for the two binding energies in Table~\ref{tab:orderbyorder} lead to such tight constraints that the resulting values of $\cE$ differ by many $\sigma$---at least in the region where the $fT_{1/2}({}^3{\rm H})$ datum is also reproduced.

The contrast when truncation errors are added to the analysis is striking; see Fig.~\subref*{fig:single_obs_constraints_with_truncation}.
In this case, the constraints due to all four observables can be satisfied simultaneously.
Note that we have fixed $Q=0.33$, $\cbar=1$, rather than marginalizing over $Q$ and $\cbar$ as we did to obtain Fig.~\ref{fig:posterior_cd_ce}.
With only one observable in the likelihood there is not enough information to determine $\cD$, $\cE$, $Q$, and $\cbar$ simultaneously.
The \NN{} LECs are also held fixed for this portion of the analysis because their effects are hardly distinguishable here.
The concordance region where all four data are simultaneously reproduced is qualitatively similar to the result obtained via MCMC sampling as in Fig.~\ref{fig:posterior_cd_ce}, though fixing $\cbar$ and $Q$ 
produces credibility intervals that are narrower than they should be, and turns tails that should be $t$ distributed back into Gaussians. 

Pairs of the triton and ${}^4$He binding energies and the $^4$He radius have conventionally been used in past optimizations of $\cD$ and $\cE$.
But Figs.~\subref*{fig:single_obs_constraints_no_truncation} and \subref*{fig:single_obs_constraints_with_truncation} make it clear that all three of these observables are correlated: they do not provide complementary constraints on the \TNF\ LECs. The triton $\beta$-decay rate---or some other non-degenerate observable---is essential to accurate estimation of $\cD$ and $\cE$~\cite{Lupu:2015pba,Epelbaum:2018ogq,Kravvaris:2020lhp,Maris:2020qne}. To make this point clear Fig.~\ref{fig:posterior_cd_ce_subsets} shows the $\cD$--$\cE$ posterior for four pairs of observables (once again with $\cbar=1$, $Q=0.33$ and fixed $\lecs_{\NN}$). 
The one-dimensional nature of the information obtained on the \TNF\ LECs under a poor choice of observable pair is most drastic for $E({}^3{\rm H})$ and $E({}^4{\rm He})$ (upper-left panel). These two binding energies are, of course, correlated: few-body universality predicts that once the three-body binding energy is known the four-body binding energy can be accurately predicted~\cite{Tjon:1975sme,Platter:2004ns,Hammer:2010kp}. 
Between them $E({}^3{\rm H})$ and $E({}^4{\rm He})$ constrain only the
combination \verifyvalue{$\cE - 0.2 \cD$}. Any information on the
individual LECs comes only from the prior, which truncates the
posterior once $|\cD| \approx \verifyvalue{5}$. The situation is
almost as bad if the $E({}^3{\rm H})$ binding energy and the $r({}^4{\rm He})$ radius are used to constrain the \TNF (upper-right panel) (cf.~the similar posterior from these two observables found in Ref.~\cite{Kravvaris:2020lhp}). 

The triton half-life constrains the value of $\cD$ well, but leaves $\cE$ essentially unconstrained~\cite{Gazit:2009PRL}. Therefore, it provides a complementary constraint, as observed in Ref.~\cite{Lupu:2015pba} (lower-left and lower-right panels), greatly reducing the range of allowed $\cD$ values. That in turn sharpens the estimate of $\cE$ because of the correlation induced through an energy or radius. 
Using the $^4$He binding energy and the triton half-life provides essentially the same information as fitting to all four observables.
Of the observables we consider, these are the two that best constrain the short-distance pieces of the \TNF.
There is little additional information added by the other two observables.
%

\section{Summary and outlook} \label{sec:summary}

The present work is part of an ongoing effort to develop, apply, and evaluate Bayesian statistical methods for effective field theories of nuclei.
Our immediate target is the estimation of the LECs $\cD$ and $\cE$ that characterize short-distance effects in the leading three-nucleon force in \chiEFT\@.
Performing this ``fit'' means finding the joint posterior distribution of these LECs given a selected set of experimental data $\genobsexpset$ and a specification of prior information, $I$, namely $\pr(\cD,\cE \given  \genobsexpset, I)$.
In this analysis, $I$ includes knowledge about the LECs as well as the \chiEFT\ truncation error model developed in Refs.~\cite{Wesolowski:2018lzj,Melendez:2019izc}.
The prior for $\cD$ and $\cE$ is chosen to be naturally sized, the $\lecs_{\NN}$ prior was determined from \NN{} scattering data up to 290\,MeV, and the \piN{} LECs were fixed to the central values from the Roy-Steiner analysis.
The resulting posterior is shown in Fig.~\ref{fig:posterior_cd_ce}.
    
We focus on how different combinations of experimental observables impact the posterior. We present results for one \chiEFT\ Hamiltonian and constrained its parameters using a set of four nuclear properties: the triton binding energy and half-life, and the ${}^4$He binding energy and charge radius.
We do not span all possible Hamiltonian regularization schemes and input properties. However, our Bayesian framework accounts for experimental and theoretical errors and  enables the identification of correlations and the direct propagation of uncertainties to observables.
Extending the results is straightforward via our open-source python package \texttt{fit3bf}~\cite{fit3bf}, which can reproduce all results shown in this work. 
    
The Bayesian strategy and the details of the statistical model are laid out in Sec.~\ref{sec:strategy}, building on previous work.
The likelihood in Eq.~\eqref{eq:likelihood} is determined by the form of the experimental and theory uncertainties to be a multivariate Gaussian. The prior information specifies \NN{} and $\pi$N LECs, as well as the uncertainties from the \NN{} fit (omitting the $\pi$N uncertainties here because we do not account for correlations with \NN{} observables).
The truncation error model for the EFT has been developed and validated elsewhere.
All assumptions are explicit and therefore testable.
    
We compute a joint posterior the LECs $\cD$, $\cE$, $\lecs_{\NN}$, and truncation error parameters $\cbar$ and $Q$.
The posterior for $\cD$ and $\cE$---unconditional on $\lecs_{\NN}$, $\cbar$, and $Q$---is obtained via marginalization.
Sampling of such an extended joint posterior is characteristic of a full Bayesian analysis.
It is made convenient and efficient here by the use of EC emulators (see Sec.~\ref{sec:emulator}).

Here are the takeaway points from this investigation:
\begin{itemize}
\item \emph{For \TNF parameter estimation, do not only use observables that are related by universality.} The triton and $\alpha$-particle binding energies and the ${}^4$He radius provide very similar constraints on $\cD$ and $\cE$ because they are related by universality; see Fig.~\subref*{fig:single_obs_constraints_with_truncation}. Consequently any pair of them only determines one linear combination of the \TNF\ LECs; see Fig.~\ref{fig:posterior_cd_ce_subsets}.
In contrast, the triton half-life provides a new constraint. When paired with the $^4$He binding energy it essentially saturates the information available from this set of observables. 
These results support the previous conclusions of Lupu \etal.~\cite{Lupu:2015pba}.
It will be interesting to make similar correlation comparisons using the three-body scattering input advocated in Refs.~\cite{Epelbaum:2018ogq,Maris:2020qne} or the information on $n\alpha$ scattering used for \TNF\ LEC estimation in Refs.~\cite{Lynn:2015jua,Kravvaris:2020lhp}.

\item \emph{The LECs $\cD$ and $\cE$ are strongly correlated.}
The contours in the joint posterior manifest a correlation of $\rho \approx 0.96$ for the \chiEFT\ Hamiltonian used in this investigation; see Fig.~\ref{fig:posterior_cd_ce}.
A similar degree of correlation was found by Baroni \etal.~\cite{Baroni:2016aa} and Kravvaris \etal.~\cite{Kravvaris:2020lhp}.
Using SCS potentials, Epelbaum \etal.~\cite{Epelbaum:2018ogq} also find strong correlation, but the orientation of the $\cD$--$\cE$ contours in that study is opposite.
Different choices of regularization scheme and scale affect the relationship between $\cD$ and $\cE$, but the details of this correlation remain to be investigated.

\item \emph{EFT truncation errors must be included for a complete quantification of uncertainties.}
Truncation errors fuzz up the constraints from individual observables, affecting the size of credibility regions in the $\cD$ and $\cE$ posterior. They do not affect the correlation.
This is evident in comparing single-observable fits in Fig.~\subref*{fig:single_obs_constraints_no_truncation} (no truncation error) to those in Fig.~\subref*{fig:single_obs_constraints_with_truncation} (including truncation error).
A consistent solution for all considered observables is only obtained when truncation errors are included;
without these errors, a simultaneous fit is problematic. 
Similar conclusions regarding the impact of truncation errors on the $\cD$--$\cE$ posterior were found using a smaller basket of $A=3$ and $A=4$ observables and a slightly different \NN{} potential in Ref.~\cite{Kravvaris:2020lhp}.

\item \emph{The impact of including \NN{} LEC uncertainties on the $\cD$--$\cE$ posterior is small. That of \piN{} LECs remains to be assessed.}
If truncation errors are included but \NN{} uncertainties are omitted, the changes in the posterior are almost undetectable. 
The \piN{} LECs were held fixed at the central values obtained in the Roy-Steiner analysis of Ref.~\cite{Siemens:2017}. Ideally the \piN{} LECs $c_1$, $c_3$, and $c_4$ would also be included in the set of parameters being sampled, so that the impact of their uncertainties on the $\cD$ and $\cE$ inference could be determined, and constraints on them from $A=3$ and $A=4$ observables assessed.
This was not feasible for the present work because the correlations between the \piN{} and the \NN{} LECs were not available. But our framework can accommodate the incorporation of \piN{} LECs in the vector $\lecs$. This is of particular interest because  those  LECs appear in the leading \chiEFT\ \TNF.

\item \emph{The EFT expansion parameter is $Q\approx 1/3$ for these observables.}
The distribution for $Q$ in Fig.~\ref{fig:hyperparameter_distributions} peaks at \verifyvalue{0.33} with a 20\% uncertainty, which is consistent with general \chiEFT\ considerations for the few-body observables used and with other estimations~\cite{Binder:2018pgl,Epelbaum:2019kcf,Maris:2020qne}.

\item \emph{\chiEFT\ provides a statistically consistent description of these few-body observables.}
The predictions of \chiEFT\ with the LEC values learned from four few-body observables reproduce these observables to within the \chiEFT\ uncertainty.
We verify this by propagating the LEC samples from MCMC sampling to the observables; the resulting posterior predictive distribution (ppd) is shown in Fig.~\ref{fig:posterior_predictive}.
We indeed used these observables in the fit, but the ppd demonstrates that \chiEFT\ can describe all four consistently---as long as truncation errors are included in the inference for $\lecs$ and thereby propagated to the ppd. 

\item \emph{Not all distributions are Gaussian.}
The joint distribution for $\cD$ and $\cE$ in  Fig.~\ref{fig:posterior_cd_ce} is best represented by a multivariate $t$ distribution. Its tails are \emph{not} well approximated by a Gaussian. 
In Appendix~\ref{app:to_a_t} we show why a $t$ distribution naturally emerges for these observables.
\end{itemize}
    
The Bayesian framework and statistical best practices we have exemplified, together with the computational capabilities enabled by EC emulators, provide a strong foundation for future work.
Full Bayesian parameter estimation and propagation of uncertainties to all calculated observables is now feasible.
Future avenues for parameter estimation with $A=3$ and $A=4$ observables include comparing  \chiEFT\ Hamiltonians with different ultraviolet regulators and with Delta degrees of freedom, 
including \piN{} LECs in the set of $\lecs$, identifying and testing complementary input observables, and applying truncation error models where the convergence pattern is correlated across observables~\cite{Melendez:2019izc,Drischler:2020yad}. 


\begin{acknowledgments}
We thank Alessandro Baroni and Rocco Schiavilla for clarifying discussions regarding their work. We thank Kyle Wendt for sharing $E_{1}^{A}$ matrix elements. SW and DRP thank Chalmers University of Technology for hospitality during the early stages of this work. 
We are also grateful for the stimulating environments at the ISNET-6 meeting ``Uncertainty Quantification at the Extremes'' in Darmstadt, at the Marcus Wallenberg Symposium ``Bayesian Inference in Subatomic Physics'' in Gothenburg, and at the INT program on ``Nuclear Structure at the Crossroads,'' each of which contributed appreciably to the content of this study.
This work was supported by the European Research Council (ERC) European Unions Horizon 2020 research and innovation programme, Grant agreement No.~758027 (AE, IS), the Swedish Research Council, Grant No.~2017-04234 (CF), National Science Foundation Award PHY–1913069 (RJF, JM) and CSSI program Award OAC-2004601 (RJF, DRP), DOE contract DE-FG02-93ER40756 (DRP), and by the NUCLEI SciDAC Collaboration under Department of Energy MSU Subcontract RC107839-OSU (RJF). Parts of the computations were enabled by resources provided by the Swedish National Infrastructure for Computing (SNIC) at Chalmers Centre for Computational Science and Engineering (C3SE), the National Supercomputer Centre (NSC) partially funded by the Swedish Research Council.
\end{acknowledgments}


\appendix

\newcommand{\linobs}{y}
\newcommand{\lindata}{\mathbf{y}}
\newcommand{\linx}{\vec{x}}
\newcommand{\linX}{X}
\newcommand{\linparams}{\vec{a}}
\newcommand{\linpmean}{\vec{\mu}}
\newcommand{\linpscale}{V}

\section{Linear models with variance estimation; or, Why things look \texorpdfstring{$t$}{t}} \label{app:to_a_t}

There are two types of distributions: those that are Gaussian and those that are not.
This is an appendix about those that are not. 

We will show that the $t$ distribution emerges as the posterior for the \TNF LECs because of two key facts. First, the observables are approximately linear in $\cD$ and $\cE$, and so at fixed $Q$ and $\cbar$ the posterior for $\cD$ and $\cE$ is Gaussian. Second, when $Q$ and $\cbar$ are estimated the tightest constraint on them comes from the variance in the theory covariance matrix in Eq.~\eqref{eq:theorycovar}.
Marginalizing over $Q$ and $\cbar$ to get the $\cD$ and $\cE$ posterior therefore corresponds to marginalizing over the variance. In linear parameter estimation problems with variance estimation the parameters are typically $t$ distributed, for the reasons we now articulate. 

Suppose that the order $k$ contributions to observables of interest $\linobs$ are linearly related to the EFT parameters $\linparams$ that appear at that order.
This is approximately true if sub-leading corrections are perturbative---as long as $k$ does not correspond to the EFT's leading order. 
In this situation the theoretical discrepancy due to truncation error, $\epsilon$, will be additive:
That is,
\begin{align}
    \linobs_k(\linx) = \linx \cdot \linparams + \epsilon.
\end{align}
If we have $N$ $O(Q^k)$ observables that we are using to extract $\linobs$ 
\begin{align}
    \lindata_k = \linX \linparams + \epsilon,
    \label{eq:OQkerrormodel}
\end{align}
where it is important to remember that the nuclear matrix elements $\linX$ that relate the LECs to the observables must be of $O(Q^k)$ if the power counting is to be valid.%
\footnote{In general there is also a contribution to $\lindata_k$ that is independent of all the $O(Q^k)$ LECs. We do not notate that here, but it can be included by defining the left-hand side of Eq.~\eqref{eq:OQkerrormodel} to be the piece of $\lindata_k$ that depends on the LECs.} 
We then write the truncation error as
\begin{align}
    \pr(\epsilon \given \cbar^2, Q) \sim \normal[0, \cbar^2 Q^{2(k+1)}] .
\end{align}
This assumes that the truncation error is the same for all observables; this assumption  can be relaxed if needed by promoting $Q$ to a matrix or including another matrix factor (say, $\genobsrefset^{\phantom{T}} \genobsrefset^T$; see~\cite{Melendez:2019izc}).

Further progress requires priors on $\linparams$ and $\cbar^2$.
We follow~\textcite{Melendez:2019izc} and place a normal-inverse-chi-squared prior on this tuple
\begin{align}
    \pr(\linparams, \cbar^2) \sim \normal\invchisq[\linpmean_0, \linpscale_0, \nu_0, \tau_0^2] ,
\end{align}
which implies that
\begin{align}
   \pr(\linparams \given \cbar^2) & \sim \normal[\linpmean_0, \cbar^2 \linpscale_0], \\
   \pr(\cbar^2) & \sim \invchisq[\nu_0, \tau_0^2].
\end{align}
The normal inverse $\chi^2$ prior is a conjugate prior and thus the posterior is the same type of distribution but with updated parameters $\linpmean$, $\linpscale$, $\nu$, $\tau^2$.
The derivation for these new parameters can be found in~\cite{Melendez:2019izc}; here we repeat the results:
\begin{align}
    \pr(\linparams \given \cbar^2, \lindata_k, Q) & \sim \normal[\linpmean, \cbar^2 \linpscale], \label{eq:Normal} \\
   \pr(\cbar^2 \given \lindata_k, Q) & \sim \invchisq[\nu, \tau^2].
\end{align}
where
\begin{align}
    \linpmean & = \linpscale {\left[\linpscale_0^{-1} \linpmean_0 + \linX^T \lindata_k / Q^{2(k+1)} \right]} \label{eq:mu} \\
    \linpscale & = \left[\linpscale_0^{-1} + \linX^T \linX / Q^{2(k+1)} \right]^{-1} \label{eq:V}\\
    \nu & = \nu_0 + N \\
    \nu\tau^2 & = \nu_0\tau_0^2 \label{eq:tau_posterior_linear} \\
    & + (\lindata_k - \linX \linpmean_0)^T [Q^{2(k+1)}\mathds{1} + \linX \linpscale_0 \linX^T]^{-1} (\lindata_k - \linX \linpmean_0). \notag
\end{align}
The limit in which the $\cbar^2$ prior is uninformative occurs when $\linpscale_0^{-1} \to 0$. Taking that limit is made easier in Eq.~\eqref{eq:tau_posterior_linear} via the Woodbury matrix identity:
\begin{align}
    [Q^{2(k+1)}\mathds{1} + \linX \linpscale_0 &\linX^T]^{-1}  = \frac{1}{Q^{2(k+1)}}\left[\mathds{1} - \frac{\linX \linpscale \linX^T}{Q^{2(k+1)}}\right] \notag \\
    & \quad\,\overset{V_0 \rightarrow \infty}{\longrightarrow} \frac{1}{Q^{2(k+1)}}[\mathds{1} - \linX(\linX^T\!\linX)^{-1}\linX^T],
\end{align}
where we have used Eq.~\eqref{eq:V} for the limit $V_0 \rightarrow \infty$ in the last line.

In the application being pursued in this work we have $\linpmean_0=0$, while $\cbar^2\linpscale_0$ is analogous to $\abar^2$ in the Gaussian prior that we impose on $\cD$ and $\cE$ in order to regulate their posteriors. Meanwhile, $\linX^T \linX$ in Eq.~\eqref{eq:V} includes terms of order $Q^{2k}$, making the second term in the square brackets of order $Q^{-2}$.
This will dominate over the first term, $\linpscale_0^{-1}$, provided that $\linpscale_0$ is natural and the values of $Q$ being marginalized over correspond to a moderately convergent EFT\@.  
The $\linpmean$ of Eq.~\eqref{eq:mu} then takes the standard form for the solution of a linear-regression problem.

Since the posterior for $\lindata_k$ is a normal distribution---albeit one with updated parameters---and the posterior for $\cbar^2$ is an inverse-chi-squared distribution,
it follows that marginalizing over $\cbar^2$ (at fixed $Q$)
yields a $t$ distribution for $\lindata_k$, see \textcite{Melendez:2019izc} for details:
\begin{align} \label{eq:lin_params_posterior}
    \pr(\linparams \given \lindata_k, Q) & \sim t_\nu[\linpmean, \tau^2 \linpscale].
\end{align}
And because this is a linear problem the posterior predictive distribution for any of the observables $y$ is also $t$:
\begin{align} \label{eq:lin_obs_posterior}
    \pr(\linobs \given \lindata_k, Q) & \sim t_\nu[\linx \cdot \linpmean, \tau^2 (\linx^T\linpscale \linx + Q^{2(k+1)})].
\end{align}
The emergence of a $t$ distribution is a standard feature in statistics problems in which the variance is unknown, and hence must be estimated from data.

This, though, does not fully explain why our results for the joint $\cD$--$\cE$ pdf follow a $t$ distribution---or at least a very good approximation to one. That result was also marginalized over $Q$.
In the uninformative limit, \ie, $\nu_0 \to 0$ and $\linpscale_0^{-1} \to 0$, the subsequent integration over $Q$ is trivial because the $Q$ dependence cancels out of Eqs.~\eqref{eq:lin_params_posterior} and~\eqref{eq:lin_obs_posterior}. Thus in this limit the result that $\linparams$ and $\linobs$ are $t$ distributed persists even after $Q$ is marginalized over. 

Insofar as our priors remain approximately uninformative, results will still be $t$ distributed.
To marginalize over $Q$ away from this limit we note that the marginalization over $\cbar^2$ and $Q$ can be formulated as a marginalization of the normal distribution~\eqref{eq:Normal} over the variance ${\cal V} \equiv \cbar^2 Q^{2(k+1)}$ and $Q$. The pdf that enters the marginalization over ${\cal V}$ is then:
\begin{align}
    \pr({\cal V}) \equiv \int \dd{Q} \dd{\cbar^2} \pr(Q,\cbar^2) \delta(\cbar^2 Q^{2k+2}-{\cal V}).
   \label{eq:Qmarginalization}
\end{align}
In our case $\pr(Q,\cbar^2)$ is the posterior for $Q$ shown in Fig.~\ref{fig:hyperparameter_distributions}. The dominant part of this distribution can be approximated by a pdf that depends only on $\cbar^2 Q^{2(k+1)}$ and not on $\cbar^2$ and $Q^{k+1}$ independently. 
Comparison of the black histogram in Fig.~\ref{fig:hyperparameter_distributions} with the red pdf for $\pr(\cbar^2,Q)$, which is an inverse $\chi^2$ distribution, suggests that 
\begin{align}
  \pr(\cbar^2 Q^{2(k+1)}) \sim \chi^{-2}(n,s^2).
\end{align}
Here $n$ and $s^2$ differ from the $\nu$ and $\tau^2$ that define the red curve and were computed using Eqs.~\eqref{eq:nu} and \eqref{eq:tausq}.
Changing variables in Eq.~\eqref{eq:Qmarginalization} from $Q$ to $u=Q^{k+1}$ we obtain
\begin{align}
        \pr({\cal V})  \propto  \frac{1}{{\cal V}^{n/2+1}} \exp\left(-\frac{n s^2}{2 {\cal V}}\right).
        \label{eq:Vresult}
\end{align}

To a good approximation the posterior for $\cD$ and $\cE$ is a Gaussian with variance ${\cal V}$.
Marginalization over ${\cal V}$ of that  posterior for $\cD$ and $\cE$ against the pdf \eqref{eq:Vresult} yields a $t$ distribution.

Therefore to the extent that EFT analyses in which $Q$ and $\cbar$ are estimated mainly constrain the variance associated with the theory uncertainty the emergence of a $t$ distribution for both the parameters and predictions is to be expected, as long as the problem is approximately linear.

\section{Optimized NN parameter values}
\label{sec:optimized_parameter_values}
The optimized values for the \NN{} LECs $\lecs_\NN$ are shown in Table~\ref{tab:nn_lec_values}. The table also includes the fixed values used for the three \piN{} LECs that enters at next-to-next-to-leading order.

\begin{table}[htpb]
  \centering
  \begin{ruledtabular}
  \setlength{\tabcolsep}{-1pt}
\begin{tabular}{Scddd}
  LEC & \multicolumn{1}{c}{\hspace*{20pt}\textrm{LO}} & \multicolumn{1}{c}{\hspace*{20pt}\textrm{NLO}} & \multicolumn{1}{c}{\hspace*{20pt}\textrm{NNLO}} \\
  \colrule \bigstrut[t]
$\widetilde{C}_{1S0}$ & -0.1115(2) & \multicolumn{1}{c}{--}& \multicolumn{1}{c}{--}\\
$\widetilde{C}_{1S0}^{np}$ & \multicolumn{1}{c}{--}& -0.1508(3) & -0.15263(8) \\
$\widetilde{C}_{1S0}^{pp}$ & \multicolumn{1}{c}{--}& -0.1504(3) & -0.15200(7) \\
$\widetilde{C}_{1S0}^{nn}$ & \multicolumn{1}{c}{--}& -0.1506(5) & -0.1523(3) \\
$\widetilde{C}_{3S1}$ & -0.0712(9) & -0.151(2) & -0.1784(8) \\
$C_{1S0}$ & \multicolumn{1}{c}{--}& ~1.458(9) & ~2.392(2) \\
$C_{3P0}$ & \multicolumn{1}{c}{--}& ~1.216(6) & ~0.999(4) \\
$C_{1P1}$ & \multicolumn{1}{c}{--}& ~0.66(3) & ~0.221(10) \\
$C_{3P1}$ & \multicolumn{1}{c}{--}& -0.239(9) & -0.974(4) \\
$C_{3S1}$ & \multicolumn{1}{c}{--}& -0.74(1) & ~0.551(5) \\
$C_{3S1-3D1}$ & \multicolumn{1}{c}{--}& ~0.19(1) & ~0.437(6) \\
$C_{3P2}$ & \multicolumn{1}{c}{--}& -0.199(2) & -0.6923(7) \\
\colrule \bigstrut[b]
$c_1$ & \multicolumn{1}{c}{--}& \multicolumn{1}{c}{--}& -0.74(2) \\
$c_3$ & \multicolumn{1}{c}{--}& \multicolumn{1}{c}{--}& -3.61(5) \\
$c_4$ & \multicolumn{1}{c}{--}& \multicolumn{1}{c}{--}& ~2.44(3) \\
\end{tabular}
\end{ruledtabular}
\caption{The \NN{} parameter values from the optimization procedure described in Sec.~\ref{sec:optimization}.
The indicated uncertainties of the \NN{} LECs (given in parentheses) correspond to the square root of the diagonal elements of the covariance matrix $\Sigma_\NN$ described in Sec.~\ref{sec:optimization}.
The \piN{} LECs $c_1$, $c_3$, and $c_4$ with corresponding uncertainties are gathered from Ref.~\cite{Siemens:2017}.
}
\label{tab:nn_lec_values}
\end{table}

\bibliography{cDcEreferences,bayesian_refs}

\end{document}